\documentclass[prx,onecolumn,nofootinbib,citeautoscript,eqsecnum,10pt,notitlepage]{revtex4-2}
\pdfoutput=1

\usepackage{dsfont}
\usepackage[caption=false]{subfig}
\usepackage{array}
\usepackage{slashed,bbold}
\usepackage{nicematrix}

\usepackage{amsmath,amssymb,bm} 
\usepackage{graphicx}

\usepackage{color} 
\usepackage[papersize={8.5in,11in}]{geometry}

\usepackage{color}
\definecolor{darkblue}{rgb}{0.,0.,0.4}
\definecolor{darkred}{rgb}{0.5,0.,0.}
\definecolor{BlueViolet}{RGB}{138,43,226}
\definecolor{SkyBlue}{RGB}{30,144,255}
\definecolor{DarkGreen}{RGB}{0,100,0}
\usepackage[pdftex,colorlinks=true,linkcolor=darkblue,citecolor=blue,urlcolor=darkred]{hyperref}

\def \be{\begin{equation}}
\def \ee{\end{equation}}
\def \nn{\nonumber \\}
\begin{document}

\title{Floquet scattering of quadratic band-touching semimetals through a time-periodic potential well}

\author{Sandip Bera}
\affiliation{Indian Institute of Technology, Kharagpur, Kharagpur 721302, India}

\author{Ipsita Mandal}
\affiliation{Institute of Nuclear Physics, Polish Academy of Sciences, 31-342 Krak\'{o}w, Poland\\
Department of Physics, Stockholm University, AlbaNova University Center,
106 91 Stockholm, Sweden}

\begin{abstract}
We consider tunneling of quasiparticles through a rectangular quantum well, subject to periodic driving. The quasiparticles are the itinerant charges in two-dimensional and three-dimensional semimetals having a quadratic band-touching (QBT) point in the Brillouin zone. In order to analyze the time-periodic Hamiltonian, we assume a non-adiabatic limit, where the Floquet theorem is applicable. By deriving the Floquet scattering matrices, we chalk out the transmission and shot noise spectra of the QBT semimetals. The spectra show Fano resonances, which we identify with the (quasi)bound states of the systems.
\end{abstract}
\maketitle

\tableofcontents


\section{Introduction}

Time-dependent driving is currently a widely used technique to influence the electronic transport properties of mesoscopic structures, for instance, a periodically driven rectangular potential well / barrier. It is important to understand the mechanisms by which the time-varying external fields affect the dynamical transport properties of these devices. Tunneling through a one-dimensional, time-modulated barrier was first theoretically considered by B\"{u}ttiker and Landauer \cite{PhysRevLett.49.1739}, where they provided an analytical framework to tackle the problem.
Since then, various effects related to oscillating potentials have been studied, which rely on the fact that an oscillating potential can transfer an incoming electron of energy $E$ (sometimes referred to as the central band), with a finite probability, to sidebands at energies $E \pm n \, \hbar\, \omega$, where $n \in \mathbb Z $ (denoting the order of the sideband) and $\omega$ is the frequency of the driving. For strongly driven systems (i.e. in the limit of high driving frequencies), a non-perturbative approach \cite{PhysRev.138.B979,PhysRevB.49.16544,PhysRevB.56.4772,
PhysRevB.60.15732} based on Floquet theory can be used, which emphasizes on the existence of sidebands of electrons exiting the potential. A sideband corresponds to an electron that has absorbed ($n>0$) or emitted ($n<0$) one or several modulation quanta $\hbar\, \omega$ \cite{PhysRevLett.49.1739}.

The Floquet scattering model consists of the incident electrons being scattered inelastically by the oscillating potential into Floquet sidebands (channels), giving rise to an infinite number of incoming and outgoing waves / channels with
quasienergies $E \pm n\,\hbar\, \omega$. This is due to the energy exchanges in units of $\hbar\, \omega $ between the incident electrons and the oscillating field. Constructing the Floquet scattering-matrix (or the S-matrix), we can derive the transmission probabilities of tunneling through the driven quantum well / barrier. In this paper, we will focus on a quantum well subject to a harmonic driving with a single frequency $\omega$.

For systems where bound states exist in the absence of a driving field (static case), even a weak driving field can cause propagating electrons at appropriate values of incident energies to undergo transitions between the spatially confined (localized) discrete bound states and the extended states in the continuum, by means of  emission / absorption of photon(s). This results in unique transmission resonances.
One of the intriguing features of these systems is that when the strength of the driving field becomes large enough, Floquet quasibound states can be created which are absent in the static system. These quasibound states appear as transmission poles in the complex energy plane \cite{PhysRevB.60.15732} and hence can also give rise to transmission resonance \footnote{These poles line  up along the real axis in the complex energy plane with the Floquet energy spacing  of $\hbar\, \omega $.}. The resonant scatterings can be related to Fano resonances \cite{Fano61,RevModPhys.82.2257}, which would appear in the transmission probabilities for both kinds of bound states, and can be observed in the shot noise spectra.

\begin{figure}[htb]
{\includegraphics[width = 0.75 \textwidth]{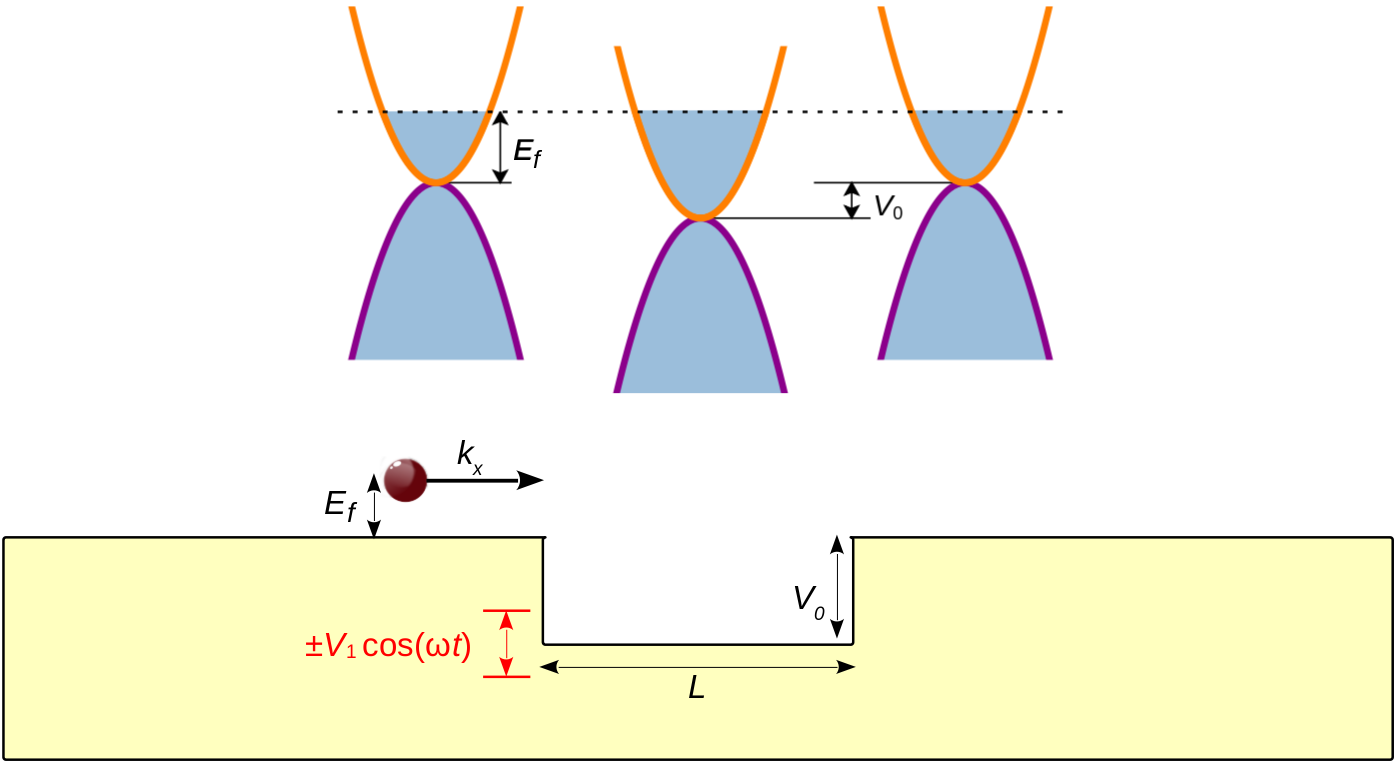}}
\caption{Tunneling through a periodically driven potential well in a QBT material. The upper panel shows the schematic diagrams of the spectrum of quasiparticles about a QBT, with respect to a potential barrier in the $x$-direction. The lower panel represents the schematic diagram of the transport across the well. The Fermi level $E_f$ is indicated by dotted lines. The blue fillings indicate occupied states. For the quantum well, $V_0$ is the depth of the static potential, and $V_1$ is the amplitude of the time-periodic drive with frequency $\omega$.
\label{figbands}}
\end{figure}

In this paper, we will compute the Fano resonances for tunneling through an oscillating electric potential well, in two-dimensional (2d) \cite{kai-sun,tsai,ips-seb} and three-dimensional (3d) \cite{Abrikosov,LABIrridate,MoonXuKimBalents} Brillouin zones harbouring quadratic band-touching (QBT) points. These semimetals are examples of multiband fermionic systems which exhibit band-crossing points in the Brillouin zone, where two or more bands cross. The set-up is schematically depicted in Fig.~\ref{figbands}. With an oscillating potential with a single frequency, both the space-inversion ($x\rightarrow -x$)  and time-reversal symmetries are preserved, and thus no d.c. current exists in the system \cite{moskalets}.

In nature, 2d QBT semimetals can be realised in checkerboard lattice at half-filling \cite{kai-sun}, Kagome  lattice at one-third-filling \cite{kai-sun}, and Lieb lattices  \cite{tsai}.
On the other hand, 3d QBT points are hosted by pyrochlore iridates ~\cite{pyro1,pyro2} $\text{A}_ 2\text{Ir}_ 2\text{O}_ 7$ (A is a lanthanide element). 3d QBT semimetals have also been realised in 3d gapless semiconductors with a sufficiently strong spin-orbit coupling \cite{Beneslavski}, and hence are relevant for materials like gray tin ($\alpha$-Sn) and mercury telluride (HgTe). They are also known as ``Luttinger semimetals''~\cite{igor16} because their low-energy fermionic degrees of freedom are captured by the Luttinger Hamiltonian of inverted band-gap semiconductors.
Tunneling of electrons for various semimetals through barriers with static electric and magnetic potentials have been studied in some recent works \cite{ips_tunnel_qbcp,ips3by2,ipsita-aritra}.


The paper is organized is follows. In Sec.~\ref{sec2dmodel}, we introduce the Hamiltonian and the scattering formalism for 2d QBT semimetals, while Sec.~\ref{sec3dmodel} deals with the 3d QBT case. We show and interpret our numerical results in Sec.~\ref{secresults}. Finally, we end with a summary and outlook in Sec.~\ref{secsum}. The details of the derivations of the scattering matrices are provided in Appendices~\ref{cal_2d} and \ref{cal_3d}.


\section{2d model}
\label{sec2dmodel}

In 2d, a particle-hole symmetric Hamiltonian harbouring a QBT point, and with $C_6$ rotational symmetry, is captured by \cite{kai-sun}:
\begin{align}
\mathcal{H}_{2d}^{kin}(p_x,p_y)&=
 \frac{\hbar^2 }{2\,\mu}
 \Big [ \, 2\, p_x\,p_y\, \sigma_x +  \left( p_y^2-p_x^2\right) \sigma_z  \,\Big ]
\end{align} 
in the momentum space (where $\mu$ is an electron's effective mass),
with eigenvalues
\begin{align}
\varepsilon_{2d}^{\pm}(p_x,p_y) = \pm  \frac{\hbar^2 \left( p_x^2 + p_y^2 \right)}{2\,\mu} \,.
\end{align}
Here the ``$+$'' and ``$-$'' signs refer to the conduction and valence bands, respectively.

We consider a quantum well of width $L$ having walls of depth $V_0$, where only the quantum well (not the adjoining regions) is subjected to a harmonic modulation of its potential with frequency
$\omega$:
\begin{align}
V(x,t)=\begin{cases} 
-V_0+V_1\cos(\omega t) &\text{ for }-L/2\le x \le L/2 \\
0 &\text{ otherwise} 
\end{cases}\,.
\end{align} 
The oscillating part of the effective potential has an amplitude $V_{1}$.
The well is assumed to be infinite and homogeneous along the $y$-direction (which means, for practical purposes, a sufficiently large width $W$), resulting in the conservation of the $k_y$-component of the momentum.
This results in the time-dependent  Schr\"{o}dinger equation:
\begin{align}
\label{schro}
    i\,\hbar\, \partial_{t} { \Psi}(x,y,t)
    =-\frac{\hbar^2}{2\,\mu} 
 \left [ \sigma_{x}
 \,\partial_{x} \,\partial_{y}  
 + \sigma_{z}\left (\partial_{y}^2 -\partial_{x}^2 \right )\right ] 
 { \Psi}(x,y,t) + \sigma_{0}\,V(x,t) \,{ \Psi}(x,y,t)  \,.
\end{align} 
The Floquet theorem \cite{PhysRev.138.B979} asserts that Eq.~\eqref{schro} admits solutions of the form $
{\Psi}(x,y,t) =
\sum \limits_{n=-\infty}^\infty e^{-i E_{n} t/\hbar}\,e^{ik_{y}y}\,\psi_n(x,t)$, where $E_{n}=E_f+n\,\hbar\,\omega$ is the  Floquet quasienergy of the $n^{\text{th}}$ order Floquet mode with $n \in \mathbb Z
$,
and $\psi_n(x,t)$ is periodic in time with periodicity $ \tau=\frac{2\,\pi}{\omega}$. Since the system is translation-invariant in the $y$-direction, we have assumed a plane-wave ansatz $e^{ik_{y}y}$ for factoring out the $y$-dependent part.
Plugging this decomposition of ${\Psi}(x,y,t)$ in Eq.~(\ref{schro}), we get:
\begin{align}
\label{schro1}
E_n\,\psi_n(x,t) =- i\,\hbar\, \partial_{t} \psi_n(x,t)
    -\frac{\hbar^2}{2\,\mu} \left [i \,k_y\, \sigma_{x}\,\partial_{x} 
- \sigma_{z}\left (k_{y}^2 +\partial_{x}^2 \right )\right ] 
 \psi_n(x,t) + \sigma_{0}\,V(x,t) \,\psi_n(x,t)  \,.
\end{align} 
Here we are looking for Floquet scattering state solutions, i.e., solutions of the Schr\"{o}dinger equation that are of the Floquet structure, with an incoming plane wave (corresponding to the conduction band) coming from $x=-\infty$ and moving along the positive $x$-axis.

In order to solve Eq.~(\ref{schro}), we need to find $\psi_n(x,t)$ piecewise in the three regions: $x<-L/2$, $-L/2 \leq x \leq L/2$, and $x>L/2$, and then impose the condition that 
the wavefunction must be continuous at the boundaries $x=\pm L/2$.
Following the formalism of Ref.~\cite{PhysRevB.60.15732,PhysRevB.84.155408}, the two-component
fermion wavefunction $ \psi_n (x, t)
=  \left (\psi_{n,1}(x,t) \quad \psi_{n,2}(x,t) \right )^T$ can be decomposed as:
 \begin{align}
 \psi_{n,1}(x,t) =
\begin{cases}  
 \gamma_{1,n}
\left \lbrace   A_{n}^{i}(t)\,e^{ik_nx} 
+A_{n}^{o}(t)\,e^{-ik_nx} \right \rbrace    & \text{ for } x< -L/2 \\ 
 \sum \limits_{m =-\infty}^\infty
\Big [ 
\gamma_{2,m} \left \lbrace  a_{m}(t)\,e^{iq_{m}x} 
+b_{m}(t)\,e^{-iq_{m}x} \right \rbrace  
J_{n-m}\big( \frac{V_{1}}{\hbar \omega}\big)
\,\Theta(E_m+ V_0) &
\\ \hspace{ 1.5 cm} +\,\gamma_{3,m} \left \lbrace   a_{m}(t)\,e^{iq_{m}x} 
+b_{m}(t)\, e^{-i q_{m}x} \right \rbrace  
J_{n-m} \big ( \frac{V_{1}}{\hbar \omega} \big) 
\, \Theta(-E_m-V_0)\Big ]
 & \text{ for } -L/2 \le x \le L/2  \\
\gamma_{1,n}
 \left \lbrace   B_{n}^{i}(t)\,e^{-ik_nx} + B_{n}^{o}(t)\,e^{ik_nx} \right \rbrace
 & \text{ for } x>L/2 
\end{cases}\,, 
\label{eqwv1}
 \end{align} 
  \begin{align}
 \psi_{n,2}(x,t)=  
 \begin{cases} 
 \gamma_{1,n}\,\frac{k_{y}}{k_n}
  \left( - A_{n}^{i} (t)\,e^{ik_nx} 
  +A_{n}^{o}(t)\,e^{-ik_nx } \right)   
 &\text{ for } x< -L/2 \\   
\sum \limits_{m =-\infty}^\infty \Big [ \gamma_{2,m} \, \frac{k_{y}}{q_{m}} 
\left \lbrace  -a_{m}(t) \,e^{iq_{m}x } 
+ b_{m}(t)\,e^{-iq_{m}x} \right \rbrace  
J_{n-m}\big ( \frac{V_{1}}{\hbar \omega}\big )
\, \Theta(E_m+V_0) &
\\ \hspace{ 1.5 cm} +\,\gamma_{3,m }\,\frac{q_{m}}{k_{y}} 
  \left \lbrace a_{m}(t)\,e^{iq_{m}x } - b_{m}(t)\,e^{-iq_{m}x} \right \rbrace
 J_{n-m} \big ( \frac{V_{1}}{\hbar \omega}\big )\, \Theta(-E_m-V_0) \Big ]
&\text{ for }   -L/2 \le x \le L/2  \\
\gamma_{1,n} \,\frac{k_{y}}{k_n}
\left \lbrace   B_{n}^{i}(t)\,e^{-ik_nx}-B_{n}^{o}(t)\,e^{ik_nx} \right \rbrace  
&  \text{ for } x>L/2
\end{cases}\,, 
\label{eqwv2}
 \end{align} 
\begin{align}
& \gamma_{1,n}=\frac{k_n}{\sqrt{k_{y}^2 +k_n^2}}\,,\quad
\gamma_{2,m}=\frac{q_{m}}{\sqrt{k_{y}^2 +q_{m}^2}}\,,\quad
\gamma_{3,m}=\frac{k_{y}}{\sqrt{k_{y}^2 +q_{m}^2}}\,,\quad
k_n=\sqrt{\frac{2\,\mu\, E_{n}}{\hbar^2}-k_{y}^2}\,,\quad
q_{m}=\sqrt{\frac{2\,\mu\, |E_{m} + V_0|}{\hbar^2}-k_{y}^2}\,,
\end{align} 
where $J_{n} (x)$ is the $n^{\text{th}}$ Bessel function of the first kind. Furthermore, $A_{n}^{i}$ and  $A_{n}^{o}$ are the amplitudes of the incoming and outgoing waves from the left, while $B_{n}^{i}$ and $B_{n}^{o}$ are those for the waves from the right, respectively. Finally, $a_{m}$ and $b_{m}$ are the  amplitudes of the wavefunction in the well region.
We note that for $E_n<0$, $ k_n$ is imaginary and represents an evanescent, meaning non-propagating mode. Such modes exist only in the neighborhood of the oscillating well and do not contribute to the current density, and hence must be omitted while computing the reflection and transmission coefficients.

The continuity of the wavefunction at the boundaries $x=\pm L/2$ gives us a matrix $ s(E_{n},E_{\tilde n})$ (see Appendix~\ref{cal_2d} for the calculational details): 
\begin{align}
s (E_{n},E_{\tilde n})=\sqrt{\frac{\text{Re}(k_n)}{\text{Re}(k_{\tilde n})}} \,
\mathcal{S}_{n {\tilde n}}\,,\quad
\begin{pmatrix} 
  { A}_{n}^{o}\\{  B}_{n}^{o}
  \end{pmatrix}   = \sum \limits_{ {\tilde n}=-\infty}^\infty \mathcal{S}_{n {\tilde n}}  
  \begin{pmatrix} 
  A_{\tilde n}^{i}\\{  B}_{\tilde n}^{i}
  \end{pmatrix}.
 \end{align} 	
Each $2\times 2$ matrix $\mathcal{S}_{n {\tilde n}}$ in the above expression encodes the probability amplitude that the electron is scattered from the $n^{\text{th}}$ order Floquet sideband to the $ {\tilde n}^{\text{th}}$ order one. The Floquet scattering matrix $s_{\alpha\beta}(E_{n},E_{\tilde n})=\sqrt{\frac{\text{Re}(k_n)}{\text{Re}(k_{\tilde n})}} \left [ \mathcal{S}_{n {\tilde n}} \right ]_{\alpha\beta}$ ($\alpha,\,\beta  \in (L, R)$)
encodes the real current flux. For a given pair of quasienergies $(E_n,\,E_{\tilde n})$, $ s(E_{n},E_{\tilde n})$
is a square matrix (whose components we denote as $s_{\alpha\beta}(E_{n},E_{\tilde n})$)
as shown below:
\begin{align}
\label{eqS2d}
s (E_{n},E_{\tilde n}) =
\begin{pmatrix} 
  s_{LL}(E_{n},E_{\tilde n}) & s_{LR}(E_{n},E_{\tilde n})\\
  s_{RL}(E_{n},E_{\tilde n}) & s_{RR}(E_{n},E_{\tilde n})\end{pmatrix} = \begin{pmatrix} 
  r_{n,{\tilde n}} & \tilde t_{n,{\tilde n}}\\t_{n, {\tilde n}} & \tilde r_{n,{\tilde n}}
  \end{pmatrix}.
\end{align}  	 
Here $r_{n, {\tilde n}}$ and $t_{n,{\tilde n}}$ are the reflection and transmission amplitudes, respectively, involving transitions of the electron from the $ {\tilde n}^{\text{th}}$ to the $n^{\text{th}}$ order Floquet channels, for modes incident from the left. On the other hand, $\tilde r_{n,{\tilde n}}$ and $\tilde t_{n,{\tilde n}}$ are the corresponding amplitudes for modes incident from the right. The unitary scattering matrix or S-matrix is obtained from $s_{\alpha\beta}(E_{n},E_{\tilde n})$ by eliminating the evanescent modes, i.e. by restricting $n$ and ${\tilde n}$ to the range $[0,\infty)$
\footnote{Note that the elements $t_{-n,0}$ and $r_{-n,0}$, with $n>0$, correspond to probability amplitudes describing an electron with incident energy $E_f$ being scattered into the evanescent mode $E_{-n}$ with energy $-n\,\hbar\,\omega$ below $E_f$.}.
It represents the quantum mechanical amplitude for an electron with energy $E_{\tilde n}$ entering the potential through lead $\beta $ to leave the well region through lead $\alpha$, after absorbing
(for $n-{\tilde n}> 0$) or emitting (for $n-{\tilde n} < 0$) $|n-{\tilde n}|\,\hbar \,\omega$ quanta of energy.

For a single electron wave incident from the left with a fixed Fermi energy $E_f$ and momentum $k_{0x}$, there is only one element, namely $\begin{pmatrix} 
  { A}_{0}^{i}\\{ B}_{0}^{i}\end{pmatrix}$, to consider for the incoming wave.
The total transmission and reflection probabilities are then given by:
\begin{align}
T=\sum \limits _{n=0} ^\infty |t_{n,0}|^2 =|s_{RL}(E_{n},E_f)|^2\,,\quad
R=\sum \limits_{n=0} ^\infty |r_{n,0}|^2=|s_{LL}(E_{n},E_f)|^2 \,,
\end{align}
respectively.

%

The components of the zero-frequency nonadiabatic pumped shot noise at low temperatures can be expressed as \cite{Dai2014,Zhu15,Zhu_2011,Zhu10,moskalets,Moskalets04}: 
\begin{align}
\label{shotnoise}
    \mathcal{N}_{\alpha \beta}(E_f) &= \frac{e^2}{h} \int_0^{\infty} dE
    \sum_{\gamma,\delta= L,R} 
\,  \sum \limits_{m,n,p=-\infty}^{\infty} 
\frac{ M_{\alpha \beta \gamma \delta}(E,E_{m},E_{n},E_{p})
\left[ f_0(E_{n}) -f_0(E_{m}) \right ]^2}{2}\,,\nn
M_{\alpha \beta \gamma \delta}(E,E_{m},E_{n},E_{p}) &=
s^*_{\alpha \gamma}(E ,E_{n}) \, s_{\alpha \delta}(E,E_{m})
\,s^*_{\beta \delta}(E_{p},E_{m})
\,s_{\beta \gamma}(E_{p},E_{n})\,,
   \end{align} 
where $f_0$ denotes the Fermi-Dirac distribution function at temperature $\mathcal{T}$. 
This measures the correlation of the current fluctuations between quasiparticle beams coming from
the $\alpha$ and $\beta$ electrodes.
Due to the current flux conservation \cite{Dai2014}, the components $\mathcal{N}_{\alpha \beta}$ have the
symmetry $\mathcal{N}_{LL}= -\mathcal{N}_{LR} = -\mathcal{N}_{LR} = \mathcal{N}_{RR}$, and hence it is sufficient to consider one of them. Here, we will pick $\mathcal{N}_{LL}$ for futher analysis.
We will also consider the differential shot noise, which is defined as the derivative of $\mathcal{N}_{LL}$ with respect to the Fermi energy.
We will take the limit $\mathcal{T} \rightarrow 0$ in our computations.



\section{3d model}
\label{sec3dmodel}

We consider a model for 3d QBT semimetals, where the low-energy bands form a four-dimensional representation of the lattice symmetry group \cite{MoonXuKimBalents}. The standard $\left(\mathbf{k} \cdot \mathbf{p} \right)$ Hamiltonian for the particle-hole symmetric system can be written by using the five $4\times 4$ Euclidean Dirac matrices $\Gamma_a$ as \cite{murakami,Herbut}:
 \begin{equation}
 \mathcal{H}_{3d}^{kin}(p_x,p_y,p_z) = \frac{\hbar^2}{2\,\mu}
 \sum_{a=1}^5 d_a(\mathbf {p}) \,  \,\Gamma_a   \,,
\label{bare}
 \end{equation}
with the anticommutator $\{ \,\Gamma_a, \,\Gamma_b \} = 2\, \delta_{ab}$.
The five anticommuting gamma-matrices can always be chosen such that three are real and two are imaginary \cite{murakami,igor12}. Here, we will use the representation such that $(\Gamma_1, \Gamma_2, \Gamma_3)$ are real, and $(\Gamma_1, \Gamma_3 ) $ are imaginary \cite{murakami}:
\begin{align}
\Gamma_1 = \sigma_3 \otimes \sigma_2 \,, \quad  \Gamma_2 = \sigma_3 \otimes \sigma_1 \,,
\quad \Gamma_3 = \sigma_2 \otimes \sigma_0 \,, \quad \Gamma_4 = \sigma_1 \otimes \sigma_0 \,,\quad
\Gamma_5 = \sigma_3 \otimes \sigma_3 \,.
\end{align}
The five functions $  d_a(\mathbf{k})$ are the real $\ell=2$ spherical harmonics given by:
\begin{align}
& d_1(\mathbf p)= -\sqrt{3}\,p_y\, p_z \, , \quad 
d_2(\mathbf k)= -\sqrt{3}\,p_x \,p_z \,, \quad
d_3(\mathbf p)= -\sqrt{3}\,p_x\, p_y  , \nonumber \\
& d_4(\mathbf p) = \frac{\sqrt{3} \left (p_y^2 - p_x^2 \right )}{2} \,, \quad
 d_5(\mathbf p) =-
\frac{2\, p_z^2 - p_x^2 -p_y^2}{2} \, .
\end{align}
The energy eigenvalues are 
\begin{align}
\varepsilon_{3d}^{\pm}(p_x,p_y,p_z) = \pm  \frac{\hbar^2 
\left( p_x^2 + p_y^2 + p_z^2 \right)}{2\,\mu} ,
\end{align}
where the ``$+$" and ``$-$" signs, as usual, refer to the conduction and valence bands. Each of these bands is doubly degenerate, and we will label it with the index $r$ or $s$. For the incident wave, we will take one of the doubly degenerate conduction bands, and label it by $r=1$ or $s=1$.
The other degenerate band will then be indicated when $r$ or $s$ takes the value $2$.

Following the same procedure as in the 2d case, we assume solutions of the form $
{\Psi}(x,y,t) =
\sum \limits_{n=-\infty}^\infty e^{-i E_{n} t/\hbar}\,e^{ik_{y}y}\,e^{ik_{z} z}\,
\psi_n(x,t)$, where $E_{n}=E_f+n\,\hbar\,\omega$ is the  Floquet quasienergy of the $n^{\text{th}}$ order Floquet mode with $n \in \mathbb Z$. The wavefunction $\psi_n(x,t)$ is now a three-component spinor which is periodic in time (with periodicity $ \tau=\frac{2\,\pi}{\omega}$).
 Since the system is translation-invariant in the $yz$-plane, we have assumed a plane-wave ansatz $e^{ik_{y}y}\, e^{ik_{z} z}$ for factoring out the $y$- and $z$-dependent parts. In this case, the piecewise decomposition of $\psi_n(x,t)$ in the three regions, $x<-L/2$, $-L/2 \leq x \leq L/2$, and $x>L/2$, can be made as:
 \begin{align}
 \psi_{n}(x,t) =
\begin{cases} 
A_{n,1}^{i}(t) \,e^{ik_{n}x}
\begin{pmatrix}
f_{in11}  \\
f_{in12}  \\
f_{in13}  \\
f_{in14}  \\
\end{pmatrix}
+A_{n,2}^{i} (t) \,e^{ik_{n}x}
\begin{pmatrix}
f_{in21}  \\
f_{in22}  \\
f_{in23}  \\
f_{in24}  \\
\end{pmatrix}
& \\
+\,A_{n,1}^{o}(t) \,e^{-ik_{n}x}
\begin{pmatrix}
f_{on11}  \\
f_{on12}  \\
f_{on13}  \\
f_{on14}  \\
\end{pmatrix}+A_{n,2}^{o}(t) \,e^{-ik_{n}x}
\begin{pmatrix}
f_{on21}  \\
f_{on22}  \\
f_{on23}  \\
f_{on24}  \\
\end{pmatrix}  
&\text{ for } x< -L/2 \\   
\sum \limits _{m=-\infty}^\infty
\Bigg[   \alpha_{m,1} (t)\,e^{iq_{m}x}
\begin{pmatrix}
\tilde f_{im11}  \\
\tilde f_{im12}  \\
\tilde f_{im13}  \\
\tilde f_{im14}  \\
\end{pmatrix}+\alpha_{m,2}(t) \,e^{iq_{m}x}
\begin{pmatrix}
\tilde f_{im21}  \\
\tilde f_{im22}  \\
\tilde f_{im23}  \\
\tilde f_{im24}  \\
\end{pmatrix} \\
+\, \beta_{m,1} (t)\,e^{-iq_{m}x}
\begin{pmatrix}
\tilde f_{om11}  \\
\tilde f_{om12}  \\
\tilde f_{om13}  \\
\tilde f_{om14}  \\
\end{pmatrix}+\beta_{m,2}(t)\, e^{-iq_{m}x}
\begin{pmatrix}
\tilde f_{om21}  \\
\tilde f_{om22}  \\
\tilde f_{om23}  \\
\tilde f_{om24}  \\
\end{pmatrix} \Bigg] 
J_{n-m}\left( \frac{V_{1}}{\hbar \omega}\right)\Theta(E_{m}+V_{0}) & \\ 
+ \sum \limits _{m=-\infty}^\infty
\Bigg[   \alpha_{m,1} (t)\,e^{iq_{m}x}
\begin{pmatrix}
\tilde g_{im11}  \\
\tilde g_{im12}  \\
\tilde g_{im13}  \\
\tilde g_{im14}  \\
\end{pmatrix}+\alpha_{m,2}(t) \,e^{iq_{m}x}
\begin{pmatrix}
\tilde g_{im21}  \\
\tilde g_{im22}  \\
\tilde g_{im23}  \\
\tilde g_{im24}  \\
\end{pmatrix}& \\
+\, \beta_{m,1} (t)\,e^{-iq_{m}x}
\begin{pmatrix}
\tilde g_{om11}  \\
\tilde g_{om12}  \\
\tilde g_{om13}  \\
\tilde g_{om14}  \\
\end{pmatrix}+\beta_{m,2}(t)\, e^{-iq_{m}x}
\begin{pmatrix}
\tilde g_{om21}  \\
\tilde g_{om22}  \\
\tilde g_{om23}  \\
\tilde g_{om24}  \\
\end{pmatrix}\Bigg ] 
J_{n-m}\left( \frac{V_{1}}{\hbar \omega}\right) \Theta(-E_{m}-V_{0}) 
&\text{ for }  -L/2 \le x \le L/2    \\
B_{n,1}^{i} (t)\,e^{ik_{n}x}
\begin{pmatrix}
f_{in11}  \\
f_{in12}  \\
f_{in13}  \\
f_{in14}  \\
\end{pmatrix}+B_{n,2}^{i} (t)\,e^{ik_{n}x}
\begin{pmatrix}
f_{in21}  \\
f_{in22}  \\
f_{in23}  \\
f_{in24}  \\
\end{pmatrix}
& \\+\,B_{n,1}^{o}(t) \,e^{-ik_{n}x}
\begin{pmatrix}
f_{on11}  \\
f_{on12}  \\
f_{on13}  \\
f_{on14}  \\
\end{pmatrix}+B_{n,2}^{o}(t)\, e^{-ik_{n}x}
\begin{pmatrix}
f_{on21}  \\
f_{on22}  \\
f_{on23}  \\
f_{on24}  \\
\end{pmatrix}
&\text{ for }  x> L/2  \\
\end{cases}\,,
\label{eq3dwave}
\end{align}
where
\begin{align}
& f_{in11}=-\frac{(k_{n}+i k_{y}) (k_{z}+\chi_{n} )}{n_{1}  (k_{n}-i k_{y})^2}\,,\quad
f_{in12}=\frac{i \,(3 k_{z}+\chi_{n} )}{\sqrt{3}\, n_{1}  (k_{n}-i k_{y})}\,,\quad
f_{in13}=-\frac{i \left(k_{n}^2+k_{y}^2-2 k_{z}^2-2 k_{z} \chi_{n} \right)}
{\sqrt{3}\, n_{1}  (k_{n}-i k_{y})^2}\,,\quad
 f_{in14}=\frac{1}{n_{1} }\,,\nonumber \\
 & f_{in21}=\frac{(k_{n}+i k_{y}) (\chi_{n} -k_{z})}{n_{2}  (k_{n}-i k_{y})^2}
 \,,\quad
 f_{in22}=-\frac{i \,(\chi_{n} -3 k_{z})}{\sqrt{3} \,n_{2}  (k_{n}-i k_{y})}\,,
 \quad
 f_{in23}=-\frac{i \left(k_{n}^2+k_{y}^2-2 k_{z}^2+2 k_{z} \chi_{n} \right)}
 {\sqrt{3} \,n_{2}  (k_{n}-i k_{y})^2}\,,
 \quad f_{in24}=\frac{1}{n_{2} }\,,
\end{align} 
\begin{align}
&f_{on11}=\frac{(k_{n}-i k_{y}) (k_{z}+\chi_{n} )}{n_{1}  (k_{n}+i k_{y})^2}\,,\quad
f_{on12}=-\frac{i \,(3 k_{z}+\chi_{n} )}
{\sqrt{3} \,n_{1}  (k_{n}+i k_{y})}\,,\quad
f_{on13}=-\frac{i \left(k_{n}^2+k_{y}^2-2 k_{z}^2-2 k_{z} \chi_{n} \right)}
{\sqrt{3}\, n_{1}  (k_{n}+i k_{y})^2}\,,\quad
f_{on14}=\frac{1}{n_{1} }\,,\nn
& f_{on21}=-\frac{(k_{n}-i k_{y}) (\chi_{n} -k_{z})}{n_{2}  (k_{n}+i k_{y})^2}
\,,\quad
f_{on22}=\frac{i \,(\chi_{n} -3 k_{z})}{\sqrt{3}\, n_{2}  (k_{n}+i k_{y})}\,,\quad
f_{on23}=-\frac{i \left(k_{n}^2+k_{y}^2-2 k_{z}^2+2 k_{z} \chi_{n} \right)}
{\sqrt{3} \,n_{2}  (k_{n}+i k_{y})^2}\,,\quad f_{on24}=\frac{1}{n_{2} }\,,
\end{align} 
\begin{align}
& \tilde f_{im11}=-\frac{(q_{m}+i k_{y}) (k_{z}+\chi_{m} )}{m_1  (q_{m}-i k_{y})^2} \,,\quad
\tilde f_{im12}=\frac{i \,(3 k_{z}+\chi_{m} )}
{\sqrt{3} \,m_1  (q_{m}-i k_{y})}\,,\quad
\tilde f_{im13}=-\frac{i \left(q_{m}^2+k_{y}^2-2 k_{z}^2-2 k_{z} \chi_{m} \right)}
{\sqrt{3} \,m_1  (q_{m}-i k_{y})^2}\,,\quad
\tilde f_{im14}=\frac{1}{m_1 }
\,,\nn
& \tilde f_{im21}=\frac{(q_{m}+i k_{y}) (\chi_{m} -k_{z})}{m_2  (q_{m}-i k_{y})^2}\,,\quad
\tilde f_{im22}=-\frac{i \,(\chi_{m} -3 k_{z})}
{\sqrt{3} \,m_2  (q_{m}-i k_{y})} \,,\quad
\tilde f_{im23}=-\frac{i \left(q_{m}^2+k_{y}^2-2 k_{z}^2+2 k_{z} \chi_{m} \right)}
{\sqrt{3} \,m_2  (q_{m}-i k_{y})^2}
\,,\quad \tilde f_{im24}=\frac{1}{ m_2 }\,,
\end{align} 
\begin{align}
&\tilde f_{om11}=\frac{(q_{m}-i k_{y}) (k_{z}+\chi_{m} )}{m_1  (q_{m}+i k_{y})^2}\,,\quad
\tilde f_{om12}=-\frac{i \,(3 k_{z}+\chi_{m} )}
{\sqrt{3} \,m_1  (q_{m}+i k_{y})}\,,\quad
\tilde f_{om13}=-\frac{i \left(q_{m}^2+k_{y}^2-2 k_{z}^2-2 k_{z} \chi_{m} \right)}
{\sqrt{3} \,m_1  (q_{m}+i k_{y})^2} \,,\quad
\tilde f_{om14}=\frac{1}{ m_1 }\,,\nn
&\tilde f_{om21}=-\frac{(q_{m}-i k_{y}) (\chi_{m} -k_{z})}{m_2  (q_{m}+i k_{y})^2}\,,\quad
\tilde f_{om22}=\frac{i \,(\chi_{m} -3 k_{z})}
{\sqrt{3} \,m_2  (q_{m}+i k_{y})}\,,\quad
\tilde f_{om23}=-\frac{i \left(q_{m}^2+k_{y}^2-2 k_{z}^2+2 k_{z} \chi_{m} \right)}
{\sqrt{3}\, m_2 
 (q_{m}+i k_{y})^2}\,,\quad
\tilde f_{om24}=\frac{1}{ m_2 }\,,
\end{align} 
\begin{align}
& n_{1}=\frac{ 2\sqrt{\frac{2\,\chi_{n}\,(\chi_{n}+k_{z} )}{3}} \, \chi_{n} }
{k_{n}^2+k_{y}^2}\,,\quad
n_{2}=\frac{ 2\sqrt{\frac{2\,\chi_{n}\,(\chi_{n}-k_{z} )}{3}} \, \chi_{n} }
{k_{n}^2+k_{y}^2}\,,\quad
 m_1=\frac{ 2\sqrt{\frac{2\,\chi_{m}\,(\chi_{m}+k_{z} )}{3}} \, \chi_{m} }
 {q_{m}^2+k_{y}^2}\,,\quad
 m_2=\frac{ 2\sqrt{\frac{2\,\chi_{m}\,(\chi_{m}-k_{z} )}{3}} \, \chi_{m} }{q_{m}^2+k_{y}^2}\,,\nn
& k_{n}=\sqrt{\frac{2\,\mu\, E_{n}}{\hbar^2}-k_{y}^2-k_{z}^2} \,,\quad
  q_{m}=\sqrt{\frac{2\,\mu \,|E_{m} + V_0|}{\hbar^2}-k_{y}^2-k_{z}^2}\,,\quad  
  \chi_{n}= \frac{2\,\mu\, E_{n}}{\hbar^2}\,,\quad
  \chi_{m}= \frac{2\,\mu \,|E_{m} + V_0|}{\hbar^2} \,.
\end{align}
We define the $\tilde g$'s in a similar manner as the $f$'s and $\tilde f$'s. However, we will not need their explicit expressions, because in our numerics we consider $E_{m}>-V_{0}$. 
  
  
The continuity of the wavefunction at the boundaries $x=\pm L/2$ now gives us the matrix $ s(E_{n},E_{\tilde n})$ (see Appendix~\ref{cal_3d} for the calculational details) as follows: 
\begin{align}
s(E_{n},E_{\tilde n})=\sqrt{\frac{\text{Re}( k_{n})}{\text{Re}( k_{\tilde n})}}\,
\mathcal{S}_{n {\tilde n}}
\,,\quad
\begin{pmatrix}
{A}_{1n}^{o}  \\
{A}_{2n}^{o}   \\
{B}_{1n}^{o} \\
{B}_{2n}^{o}   \\
\end{pmatrix}=
\sum_{\tilde n =-\infty}^\infty \mathcal{S}_{n {\tilde n}}\begin{pmatrix}
{A}_{1 {\tilde n}}^{i}  \\
{A}_{2 {\tilde n}}^{i}   \\
{B}_{1 {\tilde n}}^{i} \\
{B}_{2 {\tilde n}}^{i}   \\
\end{pmatrix}\,,
\end{align}

\begin{align}
s(E_{n},E_{\tilde n}) =
\left(\begin{NiceArray}{CC|CC}
s_{11}(E_{n},E_{\tilde n})
& s_{12}(E_{n},E_{\tilde n}) 
& s_{13}(E_{n},E_{\tilde n})
& s_{14}(E_{n},E_{\tilde n})\\
s_{21}(E_{n},E_{\tilde n})
& s_{22}(E_{n},E_{\tilde n})
& s_{23}(E_{n},E_{\tilde n})
& s_{24}(E_{n},E_{\tilde n})\\
\hline
s_{31}(E_{n},E_{\tilde n})
& s_{32}(E_{n}E_{\tilde n})
& s_{33}(E_{n},E_{\tilde n})
& s_{34}(E_{n},E_{\tilde n})\\
s_{41}(E_{n},E_{\tilde n})
& s_{42}(E_{n},E_{\tilde n})
& s_{43}(E_{n},E_{\tilde n})
& s_{44}(E_{n},E_{\tilde n})
\end{NiceArray}\right) =
\left( \begin{NiceArray}{CC|CC} 
  r^{11}_{n,{\tilde n}}
& r^{12}_{n,{\tilde n}} 
& {\tilde t}^{11}_{n,{\tilde n}}
& {\tilde t}^{12}_{n,{\tilde n}} \\
  r^{21}_{n,{\tilde n}}
& r^{22}_{n,{\tilde n}} 
& {\tilde t}^{21}_{n,{\tilde n}}
& {\tilde t}^{22}_{n,{\tilde n}} \\
\hline
  t^{11}_{n,{\tilde n}} & t^{12}_{n,{\tilde n}} 
& {\tilde r}^{11}_{n,{\tilde n}}
& {\tilde r}^{11}_{n,{\tilde n}} \\
  t^{21}_{n,{\tilde n}} & t^{22}_{n,{\tilde n}} 
& {\tilde r}^{21}_{n,{\tilde n}}
& {\tilde r}^{22}_{n,{\tilde n}}
\end{NiceArray} \right)	\,,
\end{align}

where $r^{rs}_{n,{\tilde n}}$ and $t^{rs}_{n,{\tilde n}}$ (with $r,s =1,2 $) are the reflection and transmission amplitudes, respectively, involving transitions of the electron from the $ {\tilde n}^{\text{th}}$ to the $ n^{\text{th}}$ order Floquet channels, for modes incident from the left. On the other hand, $ {\tilde r}^{rs}_{n, {\tilde n}}$ and $ {\tilde r}^{rs}_{n, {\tilde n}}$ are the corresponding amplitudes for modes incident from the right.
Due to the existence of doubly degenerate bands for the 3d QBT, we have divided the matrices into $2\times 2$ blocks. The upper left, upper right, lower left, and lower right blocks are the analogs of the $LL$, $LR$, $RL$, and $RR$ elements of the 2d case
in Eq.~\eqref{eqS2d}.
Note that the indices $r$ and $s$ distinguish between the doubly degenerate
bands that the 3d QBT has.
 
Finally, the total transmission and reflection probabilities, when the incident wave has index $r=1$, are given by:
\begin{align}
& T =\sum_{n=0}^\infty \, \sum \limits_{s=1,2}
| t^{1s}_{n,0}|^2 
=|s_{31}(E_{n}, E_f)|^2+|s_{32}(E_{n}, E_f)|^2
\,,\nn
&  R =\sum_{n=0}^\infty \,\sum \limits_{s=1,2}  |r^{1s}_{n,0}|^2
=|s_{11}(E_{n}, E_f)|^2+|s_{12}(E_{n}, E_f)|^2 \,.
\end{align}
The components of the zero-frequency nonadiabatic pumped shot noise are captured by: 
\begin{align}
\label{shotnoise3d}
\mathcal{N}_{LL}(E_f) &=N_{11} +N_{12}+N_{21}+N_{22}\,,
\quad \mathcal{N}_{LR}(E_f) =N_{13} +N_{14}+N_{23}+N_{24}\,,
\nn  \mathcal{N}_{RL}(E_f) & =N_{31} +N_{32}+N_{41}+N_{42}\,,
\quad \mathcal{N}_{RR}(E_f) =N_{33} +N_{34}+N_{43}+N_{44}\,,
\end{align} 
where
\begin{align}
   {N}_{\alpha \beta}(E_f) &= \frac{e^2}{h} \int_0^{\infty}dE \sum_{\gamma,\delta= 1}^4 
\,  \sum \limits_{m,n,p=-\infty}^{\infty} 
\frac{ M_{\alpha \beta \gamma \delta}(E,E_{m},E_{n},E_{p})
\left[ f_0(E_{n}) -f_0(E_{m}) \right ]^2}{2}\,,\nn
M_{\alpha \beta \gamma \delta}(E,E_{m},E_{n},E_{p}) &=
s^*_{\alpha \gamma}(E,E_{n}) \, s_{\alpha \delta}(E,E_{m})
\,s^*_{\beta \delta}(E_{p},E_{m})
\,s_{\beta \gamma}(E_{p},E_{n})\,.
   \end{align} 
As in the case of 2d case, here also the components $\mathcal{N}_{\alpha \beta}$ have the
symmetry $\mathcal{N}_{LL}= -\mathcal{N}_{LR} = -\mathcal{N}_{LR} = \mathcal{N}_{RR}$,
due to current flux conservation.
Again, we will pick $\mathcal{N}_{LL}$ for further analysis. We will also consider the differential shot noise, which is the derivative of $\mathcal{N}_{LL}$ with respect to the Fermi energy.
All these will be computed in the limit $\mathcal{T} \rightarrow 0$ in our computations.


\section{Numerical results and discussions}
\label{secresults}

In this section, we first show the numerical plots of the transmission coefficients and shot noise for some representative parameter values of the 2d and 3d systems (in Sec. \ref{secT2d} and \ref{secT3d}, respectively), and compare our results with other systems like graphene and pseudospin-1 Dirac-Weyl semimetals. We derive and discuss the bound state spectra in Sec.~\ref{secb}, where we also interpret their physical implications.

The minimum number of Floquet sidebands $N$ that needs to be included in our numerics is determined by the condition $N> \frac{V_{1}} {\hbar\,\omega}$, which depends on the strength of the amplitude of the oscillating part of the potential. In our plots, we set $\hbar\,\omega=4 $ meV, $L=3000$ {\AA}, $V_0=10$ meV, and $V_{1}=1$ meV, if not mentioned otherwise. We also choose $N=2$, as $V_{1} < \hbar\,\omega $.

In our numerical simulations, we have used a representative value of the effective mass, $\mu=0.001\, m_{e}$, where $m_{e}$ is the mass of a free electron. Since the effective masses are different for different materials, when trying to find the exact numbers for a given material, we need to use the appropriate value of $\mu$. For a system with $\mu=m_{f}\, m{e}$, this amounts to scaling all energy variables by the multiplicative factor $\frac {m_f} {0.001}$ in our numerics. As an example, $\mu$ in HgTe
quantum wells (which realizes 2d QBT) is around $0.03 \,m_{e}$ \cite{kvon09}, in which case we need to use the factor $\frac{0.001}{0.03}$.


\begin{figure}[]
\begin{center}
\subfloat[\label{allt}]{\includegraphics[width=0.287 \textwidth]{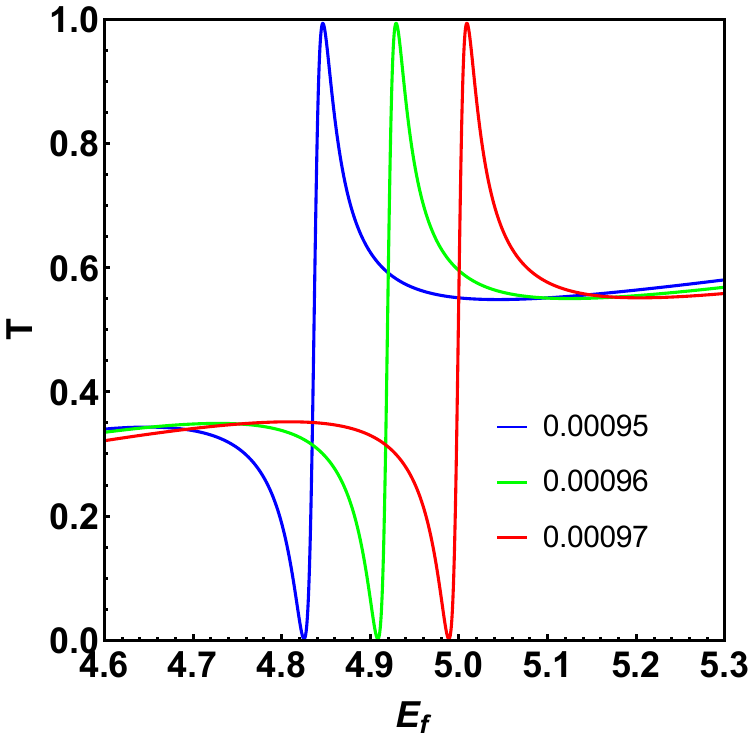}} \qquad
\subfloat[\label{noise1}]{\includegraphics[width=0.299\textwidth]{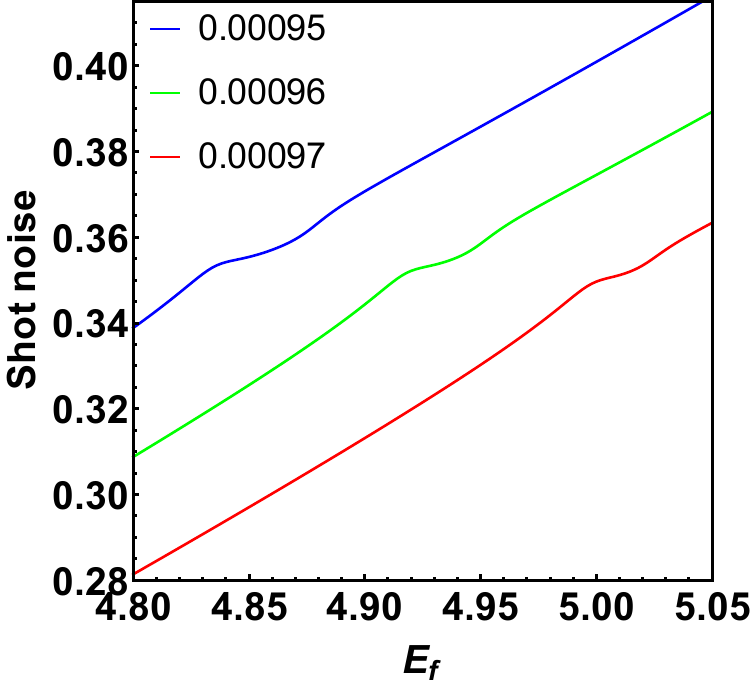}}\qquad
\subfloat[\label{derinoise}]{\includegraphics[width=0.288 \textwidth]{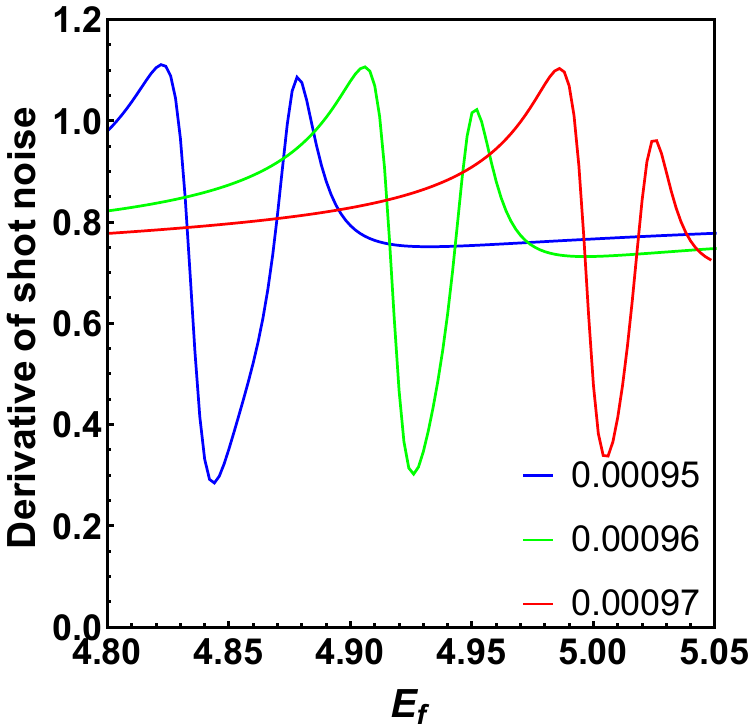}}
\caption{2d QBT: Panels (a), (b), and (c) show the total Floquet transmission coefficient ($T$), pumped shot noise (in units of $10^{-2}\,e^2\, \omega $), and the differential pumped shot noise (in units of $10^{-2}\,e^2\, h $), respectively, as functions of the energy $E_f$ (in meV) of the incident wave, for different $k_{y}$ values (in units of ${\text{\AA}}^{-1} $), as indicated in the plot-legends. The parameters used for the driven well are: $\hbar\,\omega=4 $ meV, $L=3000$ {\AA}, $V_0=10$ meV, and $V_{1}=1$ meV.
Sharp Fano resonances in $T$ can be seen, which indicate the presence of bound states in the quantum well region. Inflection points are observed in the pumped shot noise corresponding to the Fano resonances in $T$, which can be more easily identified from the plot of the derivative of the shot noise.}
	\end{center}	
\end{figure}
\begin{figure}[]
\begin{center}
\subfloat[\label{allt1}]{\includegraphics[width=0.287 \textwidth]{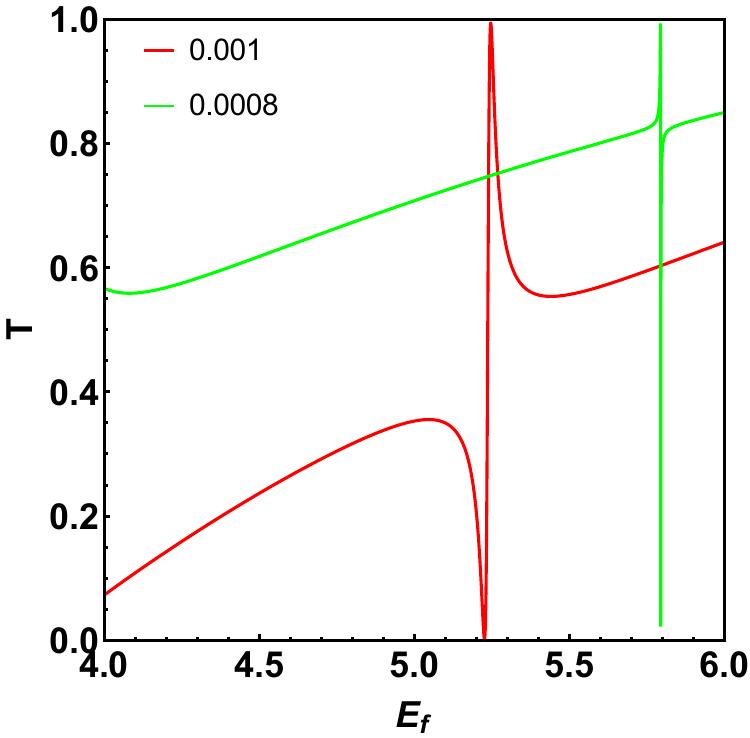}} \quad
\subfloat[\label{noise}]{\includegraphics[width=0.335\textwidth]{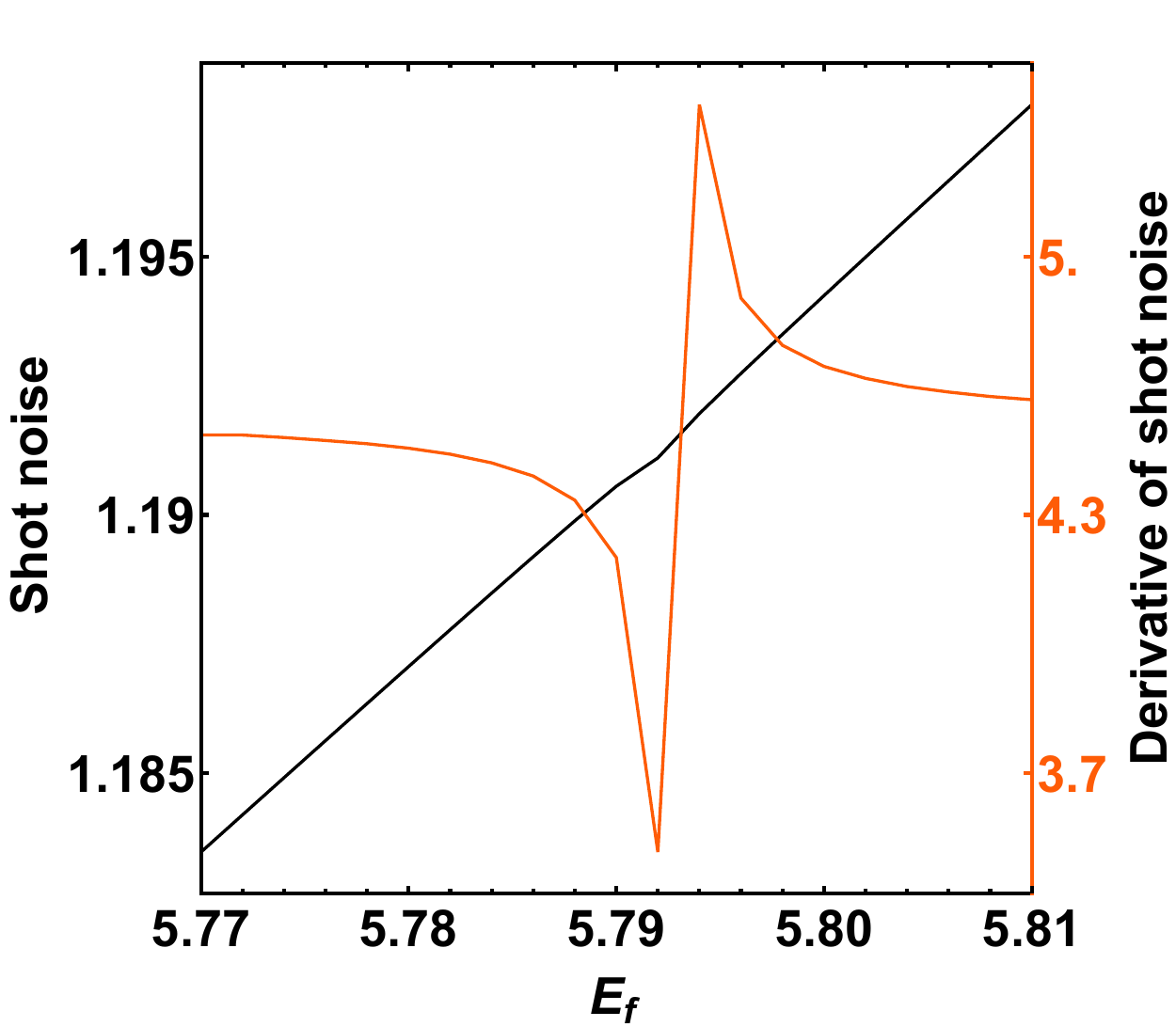}}\quad
\subfloat[\label{noise_95}]{\includegraphics[width=0.335 \textwidth]{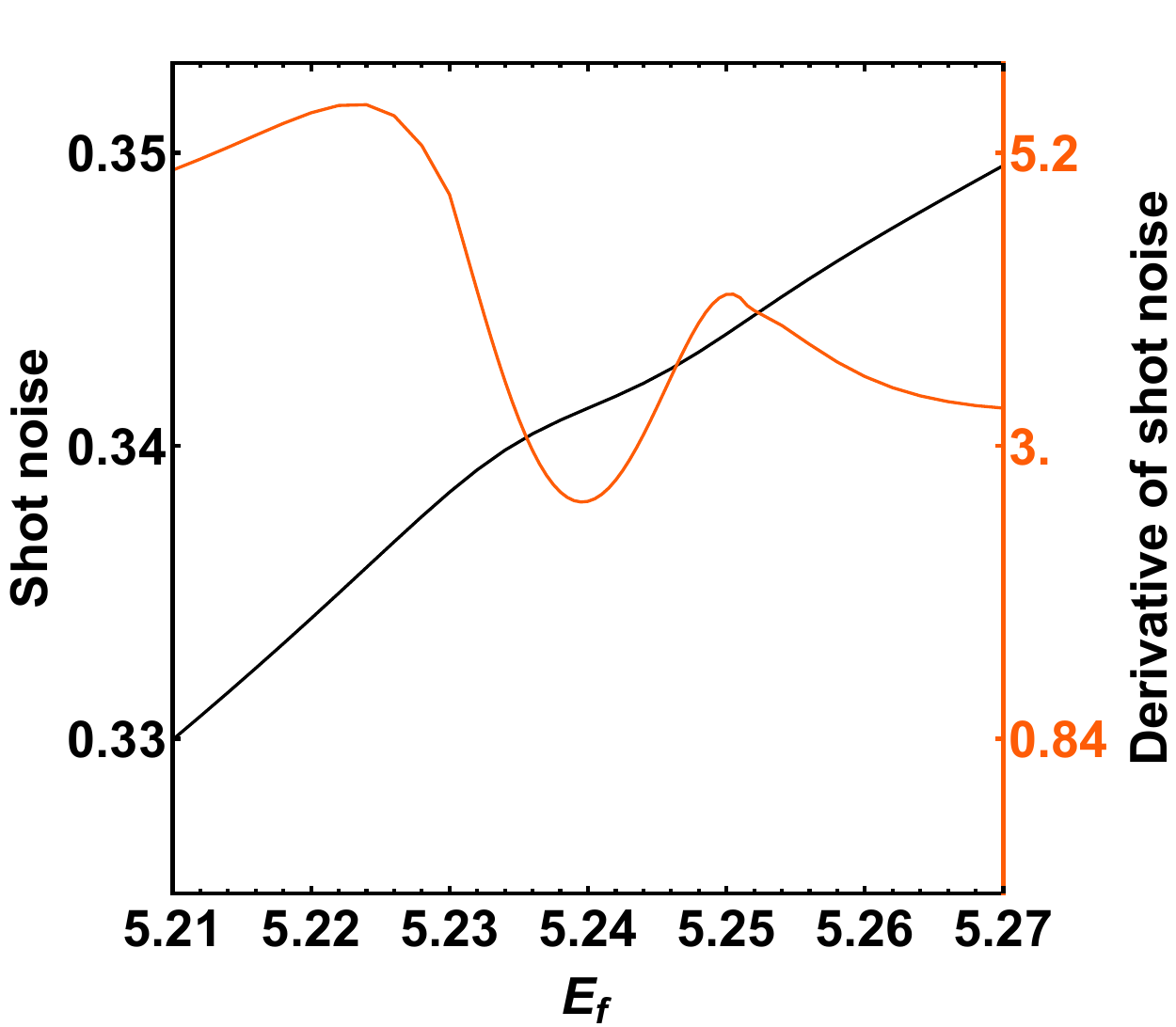}}
\caption{2d QBT: Panel (a) shows the total Floquet transmission coefficient ($T$) as a function of the energy $E_f$ (in meV) of the incident wave, for different $k_{y}$ values (in units of ${\text{\AA}}^{-1} $), as indicated in the plot-legends. Panels (b) and (c) show the pumped shot noise (in units of $10^{-2}\,e^2\, \omega $) and its derivatives (in units of $10^{-2}\,e^2\, h $) versus $E_f$ (in meV), for $k_{y}=0.0008\, {\text{\AA}}^{-1}$ and $k_{y}=0.001 \,{\text{\AA}}^{-1}$, respectively.
The parameters used for the driven well are: $\hbar\,\omega=4 $ meV, $L=3000$ {\AA}, $V_0=10$ meV, and $V_{1}=1$ meV.
Sharp Fano resonances in the plots for $T$ indicate the presence of bound states in the quantum well region. Inflection points are observed in the pumped shot noise corresponding to the Fano resonances in $T$, which can be more easily identified from the plot of the derivative of the shot noise.}
	\end{center}	
\end{figure}

\begin{figure}[]
	\begin{center}
\subfloat[\label{92}]{\includegraphics[width=0.25\textwidth]{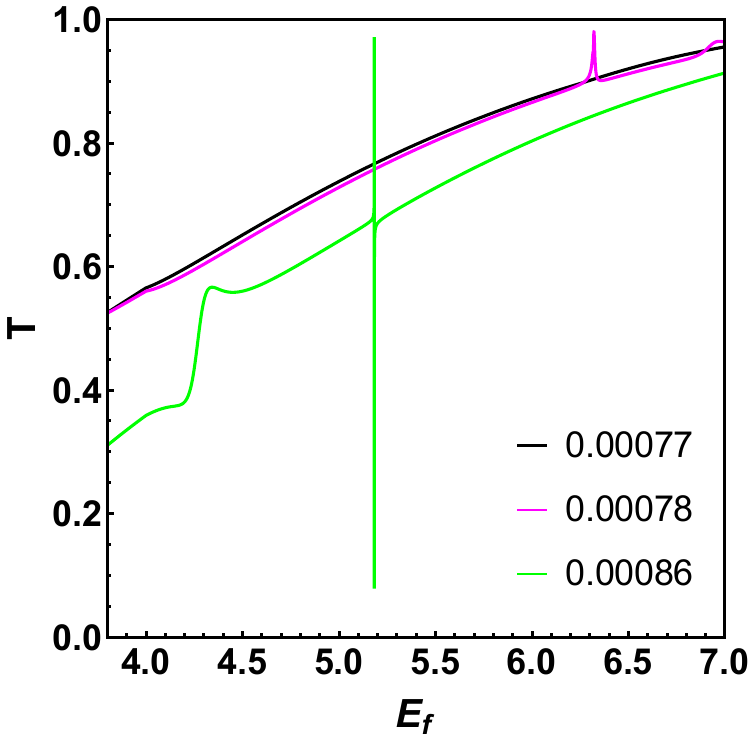}}\hspace{2 cm}
\subfloat[\label{95}]{\includegraphics[width=0.25 \textwidth]{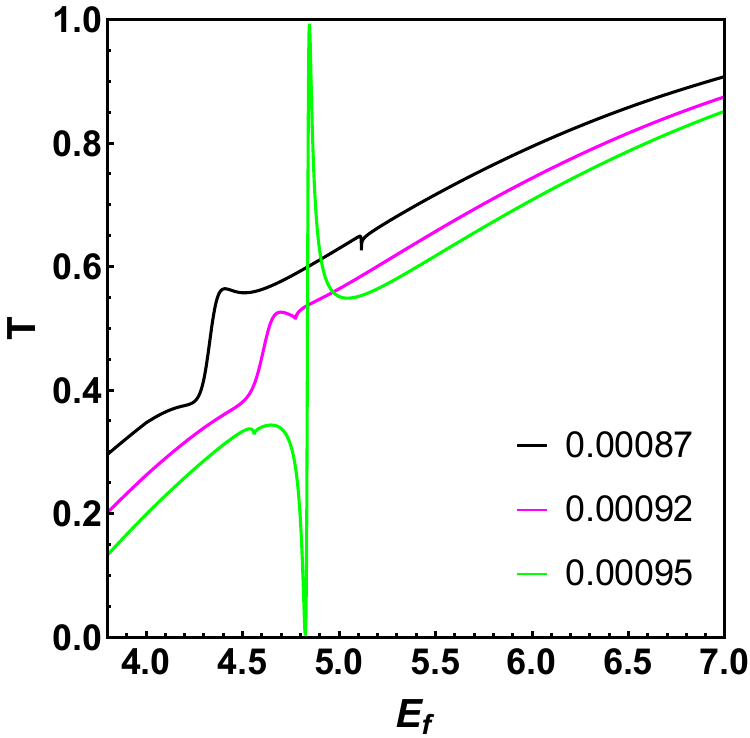}}\\ \qquad
\subfloat[\label{con_2d_tr}]{\includegraphics[width=0.3 \textwidth]{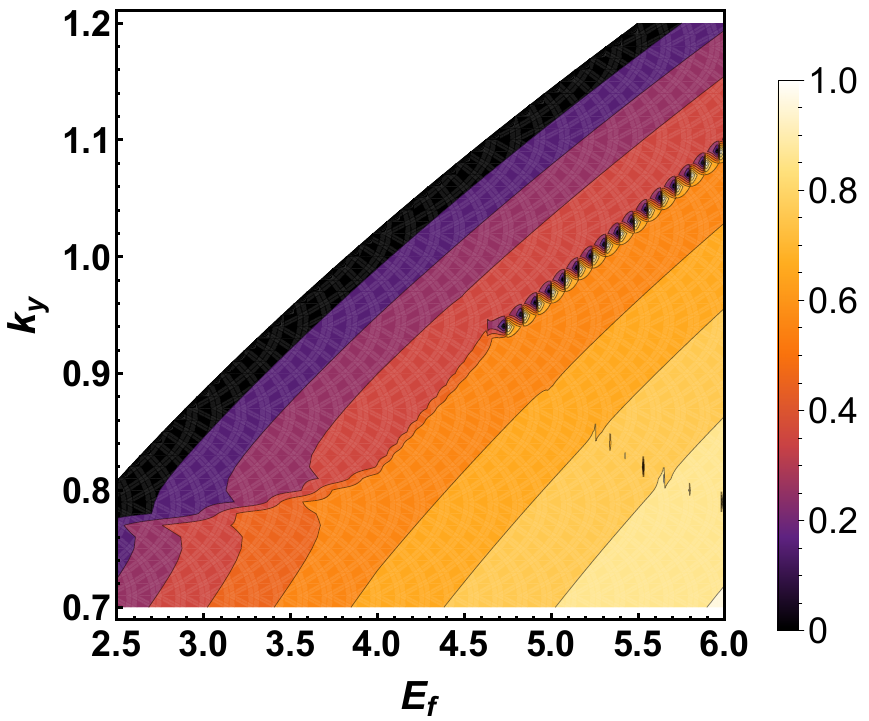}}\hspace{1.5 cm}
\subfloat[\label{con_2d_deri}]{\includegraphics[width=0.3 \textwidth]{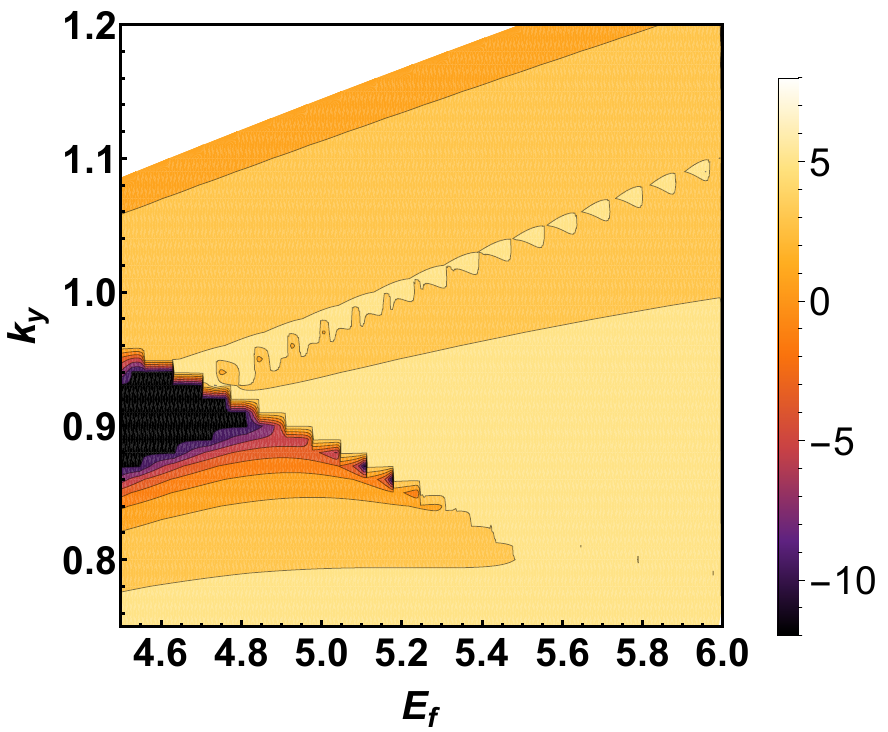}}
\caption{\label{tf_with_ky}
2d QBT: Panels (a), and (b) show the change in behaviour of the total Floquet transmission coefficient $T$ as a function of $E_f$ (in meV), as we gradually tune $k_{y}$ to different values (in units of ${\text{\AA}}^{-1}$), as shown in the plot-legends.
Panels (c) and (d) show the contour-plots of $T$ and the derivative of shot noise, respectively, in the $E_{f}$-$ k_{y}$ plane (with $E_f$ in meV and $k_{y}$ in $10^{-3}\,{\text{\AA}}^{-1} $). The latter is represented by a color code with the unit $10^{-2}e^2\, h$.
For these contour-plots, $k_y$ is in units of $10^{-3} \,{\text{\AA}}^{-1}$.
For all the panels, the parameter values of the driven well are kept fixed at $\hbar\,\omega=4 $ meV, $L=3000$ {\AA}, $V_0=10$ meV, and $V_{1}=1$ meV. 
}
\end{center}	
\end{figure}

\subsection{2d model}
\label{secT2d}

First, let us discuss the transport features of the 2d QBT.
The numerical results for the transmission coefficient $T$ as a function of the $E_f$ are shown in Figs.~\ref{allt}, \ref{allt1}, \ref{92}, and \ref{95}. Fano resonances are observed when the first order Floquet sidebands overlap with a bound state within the quantum well. The resonances that we find have asymmetric patterns, just like the cases for an electron gas and graphene \cite{PhysRevB.60.15732,Zhu15}. 
Analogous to the features for a free electron gas, the asymmetric pattern has a sharp dip is followed by a peak, which is opposite to the asymmetric pattern (namely, a perfect transmission followed by a total reflection) found for graphene. We note that this asymmmetry is in contrast with the symmetric $T$ found in systems like pseudospin-1 Dirac-Weyl systems \cite{Zhu17}, which is connected with the parity structure of the components of the wavefunction of the bound states.
We also show the corresponding shot noise ($\mathcal{N}_{LL}$) and its derivative (see
Figs.~\ref{noise1}, \ref{derinoise}, \ref{noise}, and \ref{noise_95}). 
Inflection points are observed in $\mathcal{N}_{LL}$ corresponding to the Fano resonances in the $T$, which can be more easily identified from the plot of its derivative.
We have also included representative contour-plots of $T$ and the derivative of shot noise in the $E_{f}$-$ k_{y}$ plane (see Figs.~\ref{con_2d_tr} and \ref{con_2d_deri}).

We observe two kinds of trends in the transmission features:
 \begin{enumerate}
 
 \item{Type $1$:
 Fig.~\ref{allt} shows that the value of $E_f$ at which the Fano resonance occurs (let us call this the Fano resonance point (FRP)) increase with increasing $k_{y}$, which is opposite to the trend seen in the results for graphene \cite{Zhu15} (in graphene, the $E_f$ value for FRP decreases with increasing $k_{y}$). The bandwidth of the resonance curves are almost constant with $k_{y}$. We have checked this feature for higher values of $k_{y}$ (up to $0.0015\, {\text{\AA}}^{-1}$).
Also, in graphene, the asymmetric Fano resonace pattern has a peak followed by a dip. In 2d QBT,
the opposite pattern is observed, namely, a dip is followed a peak. This is also reflected in
the shot noise and its derivative, as shown in Figs.~\ref{noise1} and  \ref{derinoise}.
} 
 
 \item{Type $2$:
 Fig.~\ref{allt1} shows the value of $E_f$ for FRP decrease with increasing $k_{y}$, which is  consistent with the features seen in graphene \cite{Zhu15}.  But, the value and sequence of change in the bandwidth of the FRP with changing $k_{y}$ are not the same as what is seen in graphene. The resonance peak for $k_{y}=0.0008 \, {\text{\AA}}^{-1}$ is very sharp (small bandwidth) compared to the $k_{y}=0.001 \, {\text{\AA}}^{-1}$ case. The nature of the shot noise and its derivative in Fig.~\ref{noise} also differ from the curves in type $1$, as the inflection region is extremely narrow and the derivative has a very sharp peak at FRP for $k_{y}=0.0008 \, {\text{\AA}}^{-1}$. Fig.~\ref{noise_95} depicts a shot noise pattern similar to the curves in type $1$.
} 

\end{enumerate}

Hence, based on the nature of the curves we conclude that two different types (type $1$ and type $2$) of Fano resonances show up depending on the parameter regimes. The gradual change in the nature of the FRPs with  increasing $k_{y}$ as we change its value from $ 0.00077{\text{\AA}}^{-1}$ to $0.00095\, {\text{\AA}}^{-1} $ is shown in Fig.~\ref{tf_with_ky}. Firstly, we see that an FRP with a relatively small peak, but broad bandwidth, appears above $k_y=0.00077{\text{\AA}}^{-1}$, and it behaves as a type $1$ peak (as the peak position shifts to higher values of $E_f$ with increasing $k_y$).
Its bandwidth decreases as $k_y$ increases. Secondly, a sharp (very narrow bandwidth) type $2$ FRP also appears above $k_y=0.00077{\text{\AA}}^{-1}$, whose peak position shifts to lower values of $E_f$ with increasing $k_y$. We note that both types of FRPs disappear in the limit $k_{y} \le 0.00077 \,{\text{\AA}}^{-1} $.


\begin{figure}[]
\begin{center}
\subfloat[\label{15t}]{\includegraphics[width=0.25 \textwidth]{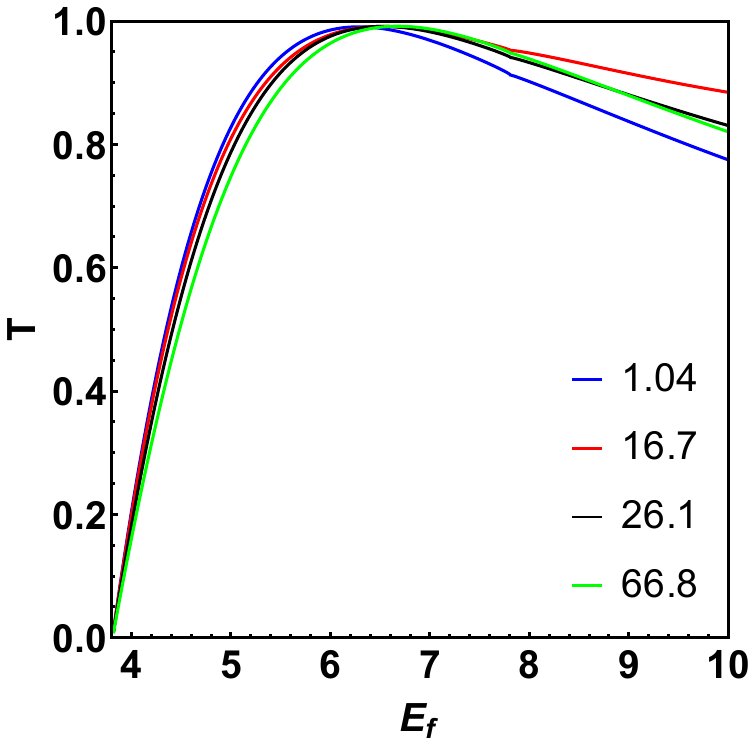}}\hspace{0.75 cm}\,
\subfloat[\label{5t}]{\includegraphics[width=0.25 \textwidth]{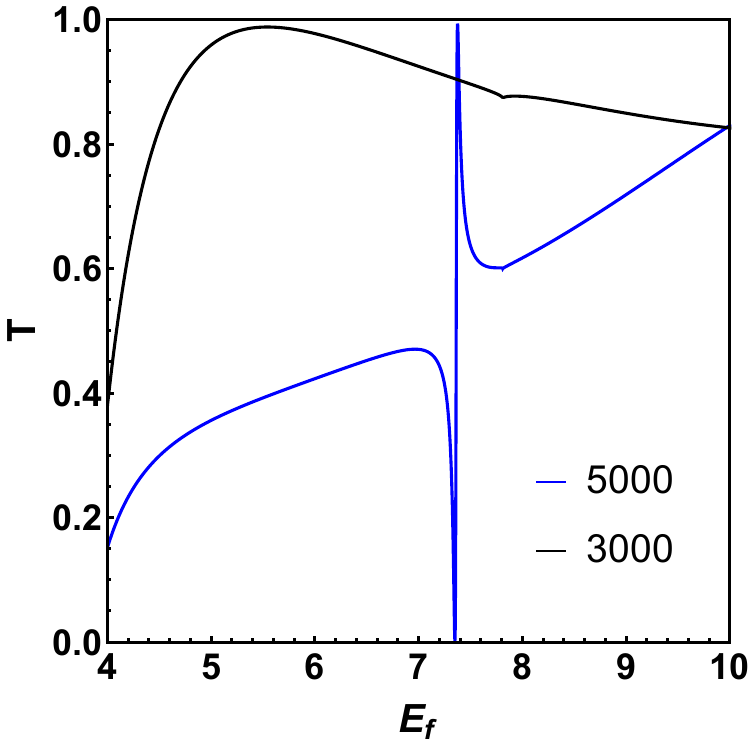}}\hspace{0.75 cm}\,
\subfloat[\label{20t}]{\includegraphics[width=0.25 \textwidth]{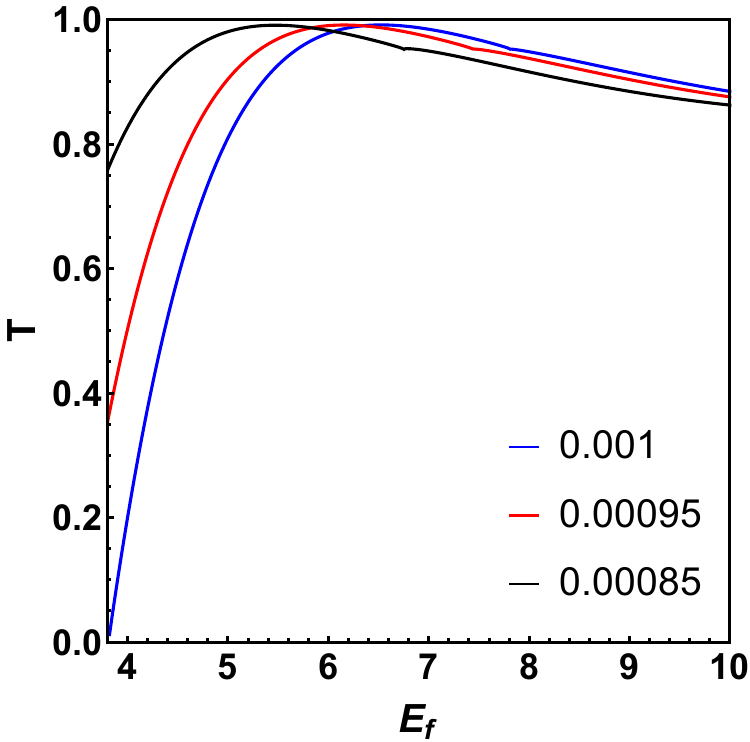}}
\caption{\label{tf_v0}
2d QBT: Panel (a) shows $T$ versus $E_f$ (in meV) for various values of $V_{0}$ (in meV), as indicated in the plot-legends, at $L=3000$ {\AA} and $k_y=0.001\, {\text{\AA}}^{-1} $.
Panel (b) shows $T$ versus $E_f$ (in meV) at $V_0=16.7$ meV and $k_y=0.001\, {\text{\AA}}^{-1} $, for  $L=3000$ {\AA} and $L=5000$ {\AA}, respectively.  
Panel (c) shows $T$ versus $E_f$ (in meV) for various values of $k_y$ (in ${\text{\AA}}^{-1} $), as indicated in the plot-legends, at $L=3000$ {\AA} and $V_{0}=16.7$ meV. The remaining parameter values are kept fixed at $\hbar\,\omega=4 $ meV, and $V_{1}=1$ meV for all the panels.}		
\end{center}	
\end{figure}

From the relations shown in Appendix~\ref{cal_2d}, we find that the matrix components of the $M_{1s}^{\pm}$ and $M_{2s}^{\pm}$ matrices in Eq.~\eqref{30} depend on the phase factors $e^{\pm \frac{i \, q_{m} \,L}{2}}$ (where $q_{m}=\sqrt{\frac{2\,\mu\, |E_{m} + V_0|}{\hbar^2}-k_{y}^2}$).  Clearly, for $ \frac{2 \,\mu\, E_{m}}{\hbar^2}=k_{y}^2$ , the phase factor becomes $ e^{ i\,\alpha} $, where $\alpha = \sqrt{\frac{2 \mu V_{0}}{\hbar^2}} \frac{L}{2}$. From our simulations, we find that FRPs are absent if $\sin{\alpha}=0$ or
$\tan{\alpha} = 1$, i.e $V_{0}= \frac{\hbar^2}{2 \mu} \left (\frac{2 \,m_{3}\,\pi}{L} \right )^2$ or $\frac{\hbar^2}{2 \mu}\left (\frac{ m_{4}\,\pi}{2\, L} \right )^2$ (where $m_{3}=1,2,3, \cdots$ and $m_{4}=1,5,9, \cdots $). Hence for $L=3000$ {\AA} and $k_y=0.001\, {\text{\AA}}^{-1} $, the FRPs are absent at $V_{0}=1.04,\, 16.71,\, 26.11,\,66.85, \cdots$ meV, which is shown in Fig.~\ref{15t}. 
On the other hand, FRPs are present if we change $L$ to $5000$ {\AA}, as seen in Fig.~\ref{5t}. Fig.~\ref{20t} shows that this phenomenon is completely independent of $k_{y}$. Although Ref.~\cite{Zhu15} does not discuss this aspect, our simulations for graphene prove that same features exist also in graphene. Hence, this seems to be a system-independent feature.

The nature of the FRPs are caused by the bound state spectra of the system. Hence, we will point out the reasons for the differences in behaviour of the 2d QBT FRPs from those in systems like graphene in Sec.~\ref{secb}, where we plot and discuss the bound state spectra.

\begin{figure}[]
\subfloat[\label{allt_3d}]
{\includegraphics[width=0.225  \textwidth]{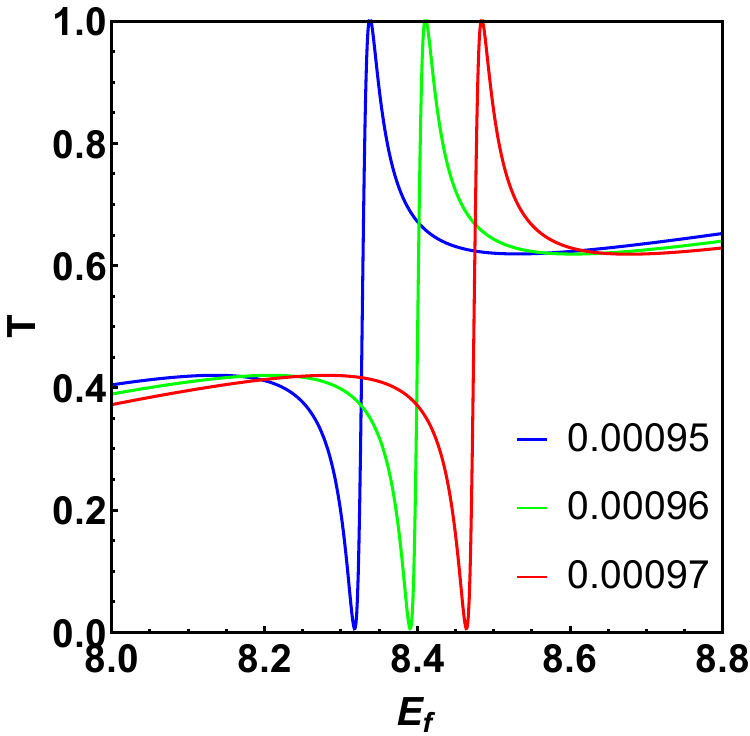}} \,
\subfloat[\label{noise3d_1}]
{\includegraphics[width=0.235 \textwidth]{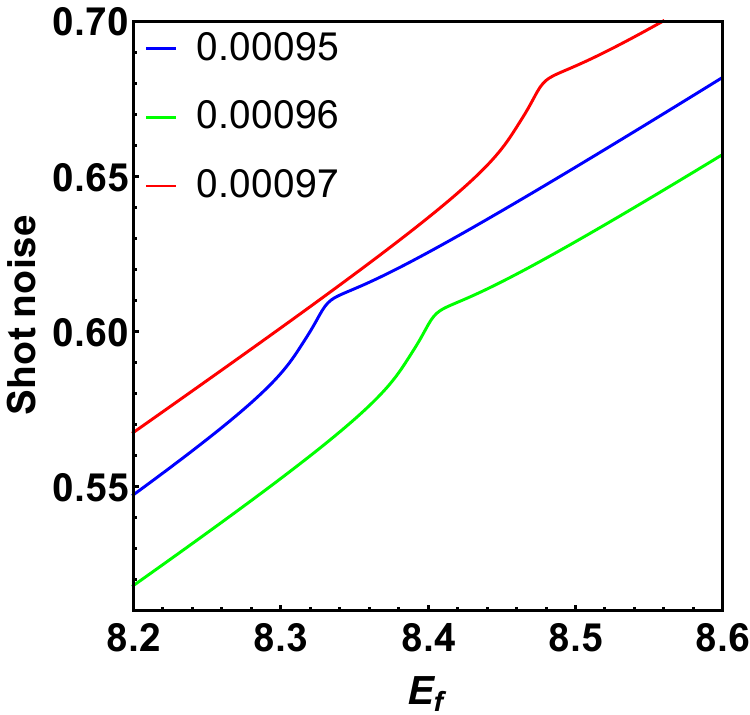}}\,
\subfloat[\label{derinoise3d_1}]
{\includegraphics[width=0.225 \textwidth]{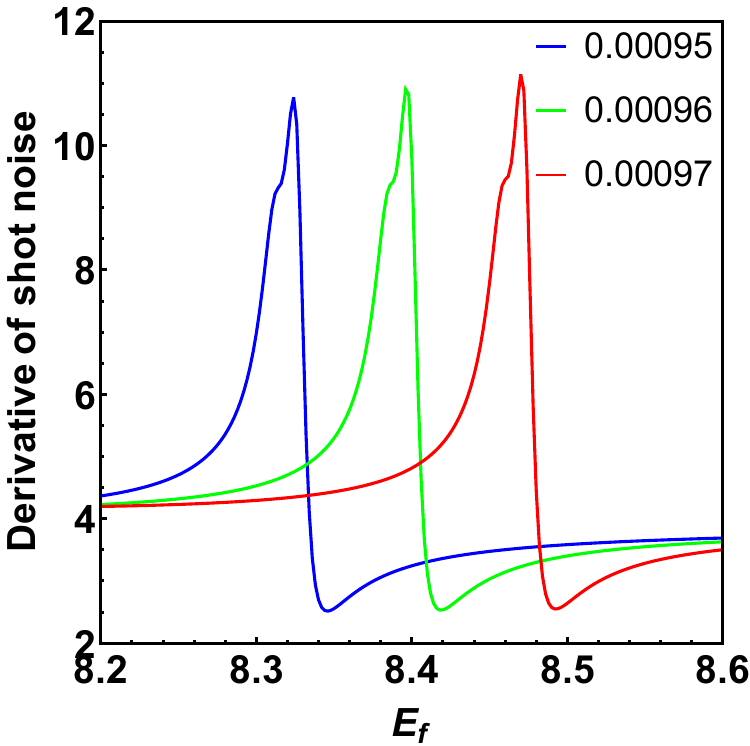}}\,
\subfloat[\label{con_3d}]
{\includegraphics[width=0.273 \textwidth]{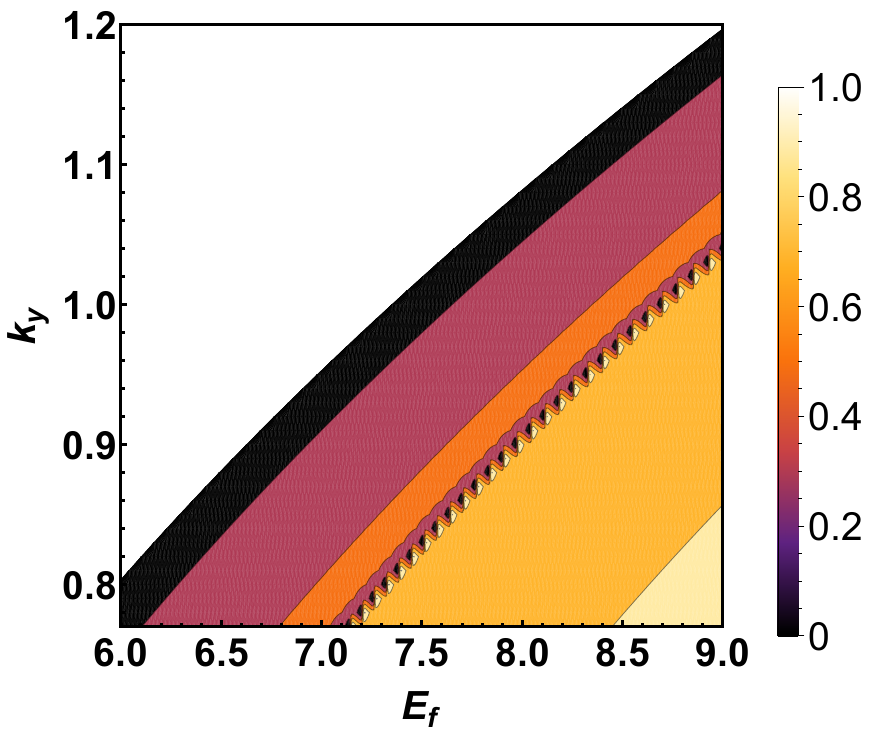}}
\caption{\label{large3d}
3d QBT: Panels (a), (b), and (c) show the total Floquet transmission coefficient ($T$), pumped shot noise (in units of $10^{-2}\,e^2\, \omega $), and the differential pumped shot noise (in units of $10^{-2}\,e^2\, h $), respectively, as functions of the energy $E_f$ (in meV) of the incident wave, for different $k_{y}$ values (in units of ${\text{\AA}}^{-1}$), as indicated in the plot-legends. The $k_{z}$ value is kept fixed at $0.00095 \,{\text{\AA}}^{-1}$.
Sharp Fano resonances in $T$ can be seen, which indicate the presence of bound states in the quantum well region. Inflection points are observed in the pumped shot noise corresponding to the Fano resonances in $T$, which can be more easily identified from the plot of the derivative of the shot noise. Panel (d) shows the contour-plot of $T$ in the $E_f$-$k_{y}$ plane (with $E_f$ in meV and $k_{y}$ in $10^{-3}\,{\text{\AA}}^{-1} $), for $k_z=0.00095 \,{\text{\AA}}^{-1}$.
For all the panels, the parameters used for the driven well are: $\hbar\,\omega=4 $ meV, $L=3000$ {\AA}, $V_0=10$ meV, and $V_{1}=1$ meV.}
\end{figure}

\subsection{3d model}
\label{secT3d}

\begin{figure}[]
\begin{center}
\subfloat[\label{t_3d2}]
{\includegraphics[width= 0.23   \textwidth]{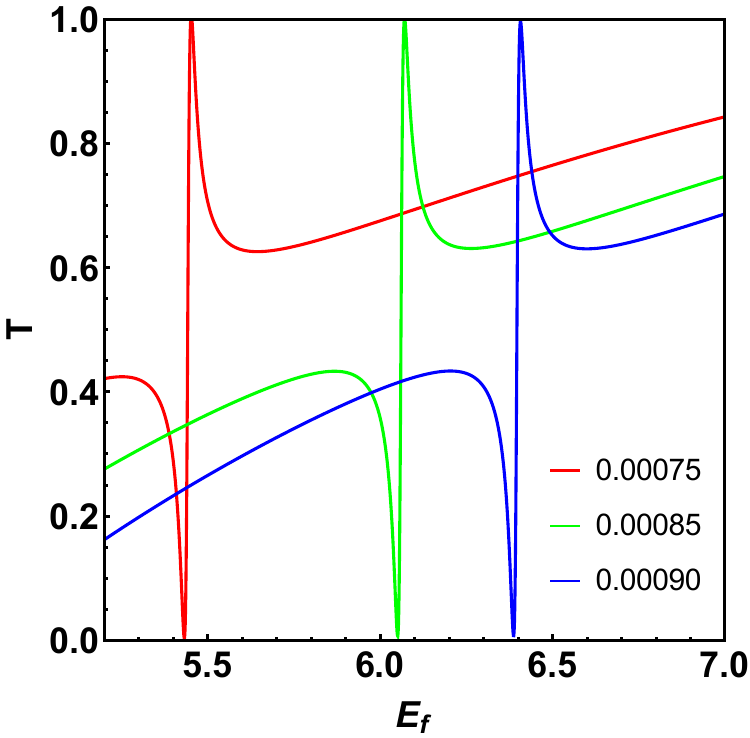}} \,
\subfloat[\label{noise3d_2}]
{\includegraphics[width= 0.233  \textwidth]{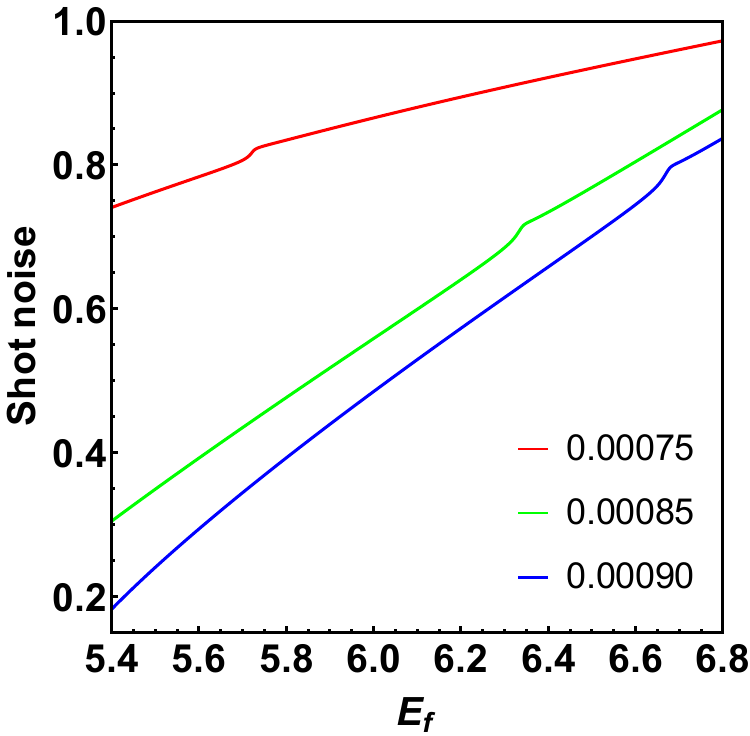}}\,
\subfloat[\label{derinoise3d_2}]
{\includegraphics[width=0.23  \textwidth]{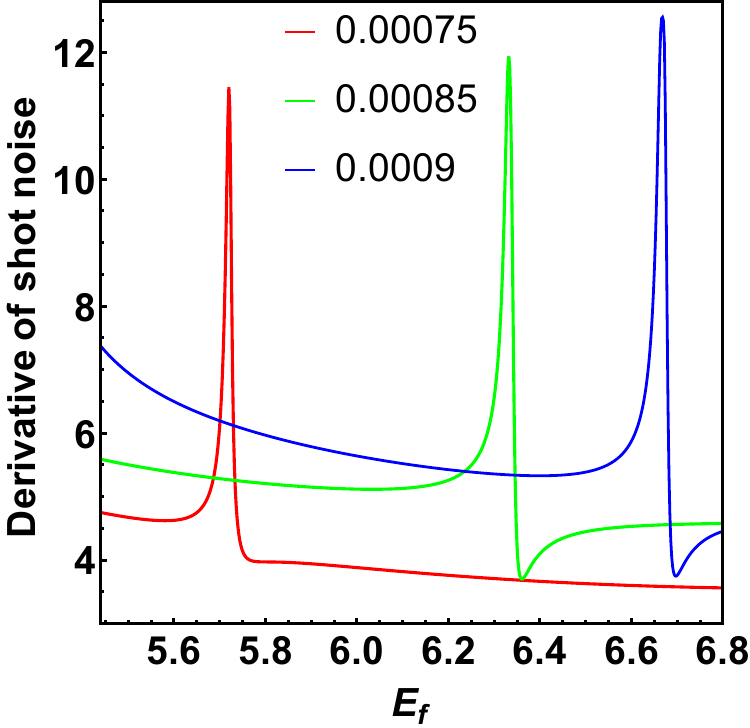}}\,
\subfloat[\label{con_3d_75}]
{\includegraphics[width=0.27 \textwidth]{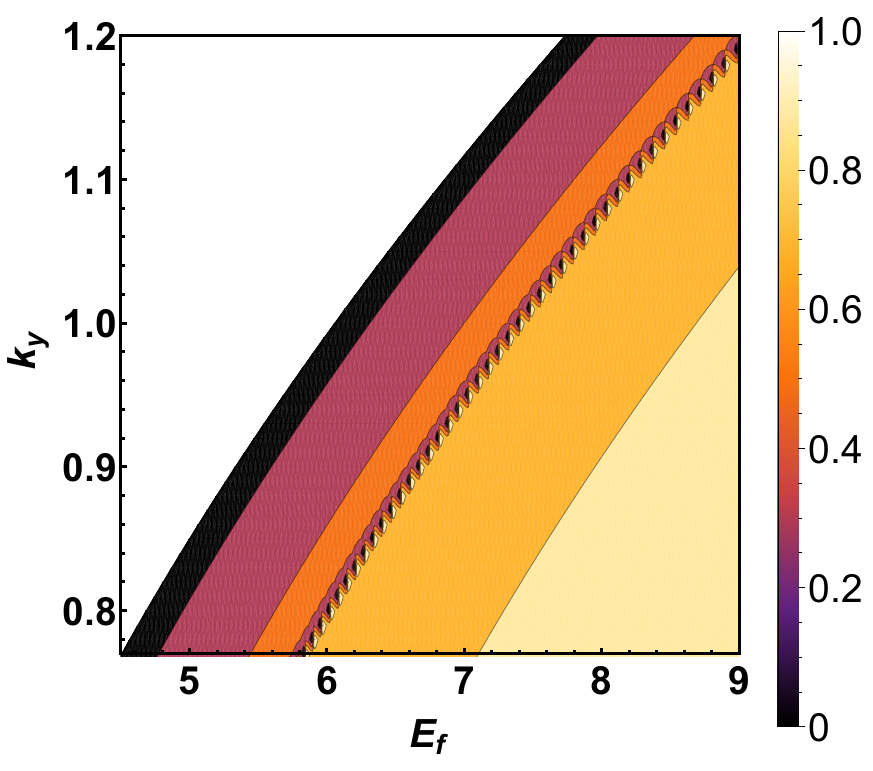}}
\caption{\label{large3d2}
3d QBT: Panels (a), (b), and (c) show the total Floquet transmission coefficient ($T$), pumped shot noise (in units of $10^{-2}\,e^2\, \omega $), and the differential pumped shot noise (in units of $10^{-2}\,e^2\, h $), respectively, as functions of the energy $E_f$ (in meV) of the incident wave, for different $k_{y}$ values (in units of ${\text{\AA}}^{-1}$), as indicated in the plot-legends. The $k_{z}$ value is kept fixed at $0.00075 \,{\text{\AA}}^{-1}$.
Sharp Fano resonances in $T$ can be seen, which indicate the presence of bound states in the quantum well region. Inflection points are observed in the pumped shot noise corresponding to the Fano resonances in $T$, which can be more easily identified from the plot of the derivative of the shot noise. Panel (d) shows the contour-plot of $T$ in the $E_f$-$k_{y}$ plane (with $E_f$ in meV and $k_{y}$ in $10^{-3}\,{\text{\AA}}^{-1} $), for $k_z=0.00075 \,{\text{\AA}}^{-1}$. For all the panels, the parameters used for the driven well are: $\hbar\,\omega=4 $ meV, $L=3000$ {\AA}, $V_0=10$ meV, and $V_{1}=1$ meV.
}
\end{center}	
\end{figure}


\begin{figure}[]
	\begin{center}
\subfloat[\label{3_5t}]{\includegraphics[width=0.23 \textwidth]{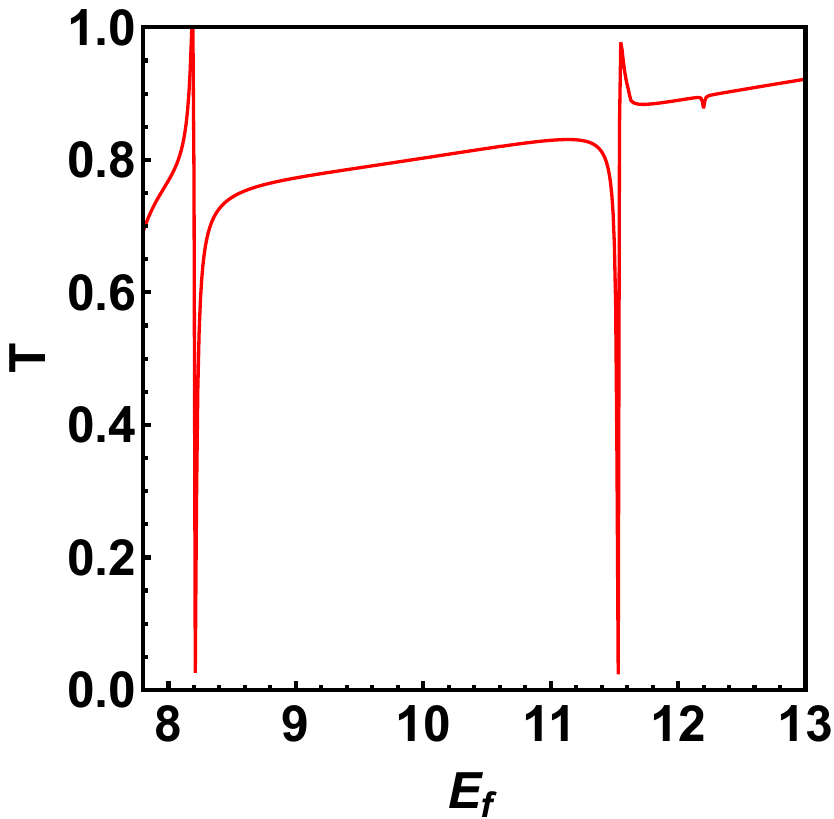}}\,
\subfloat[\label{3_10t}]{\includegraphics[width=0.23 \textwidth]{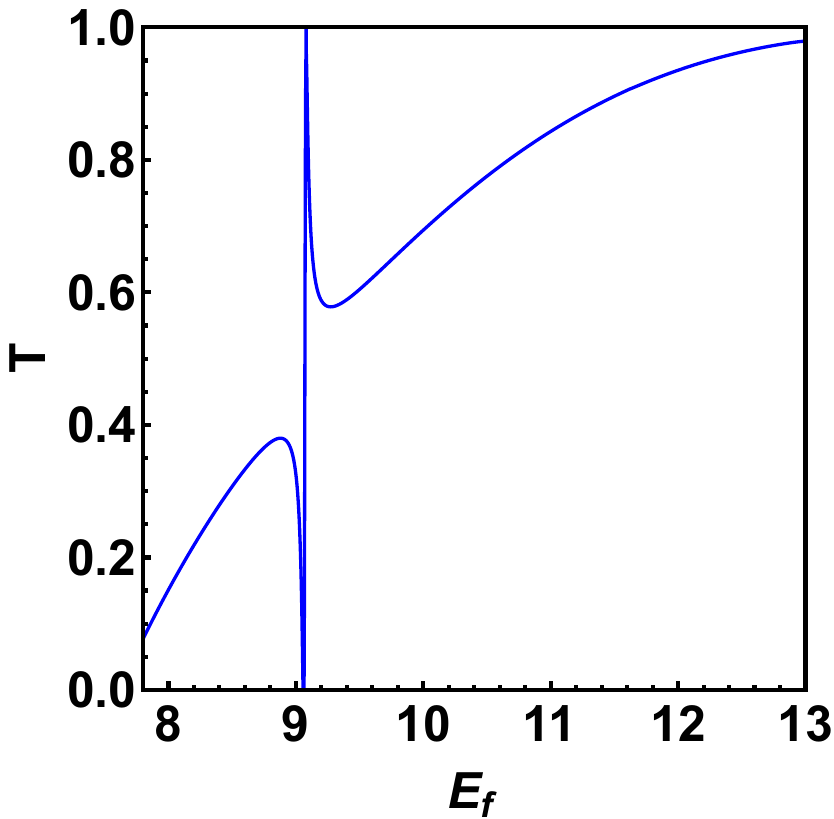}}\,
\subfloat[\label{3_15t}]{\includegraphics[width=0.23 \textwidth]{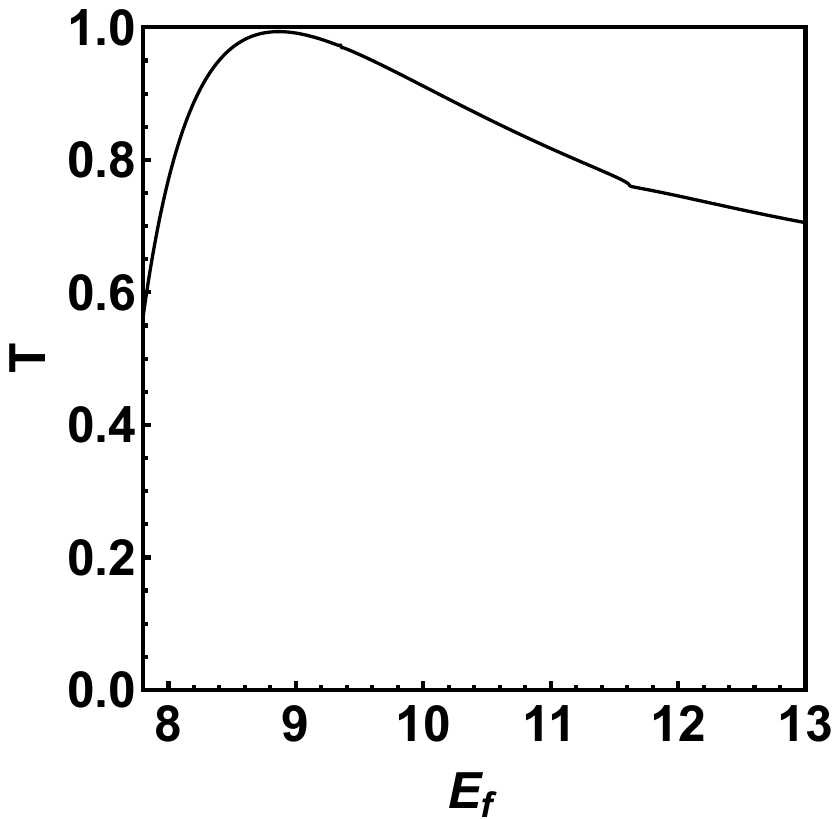}}\,
\subfloat[\label{3_20t}]{\includegraphics[width=0.23 \textwidth]{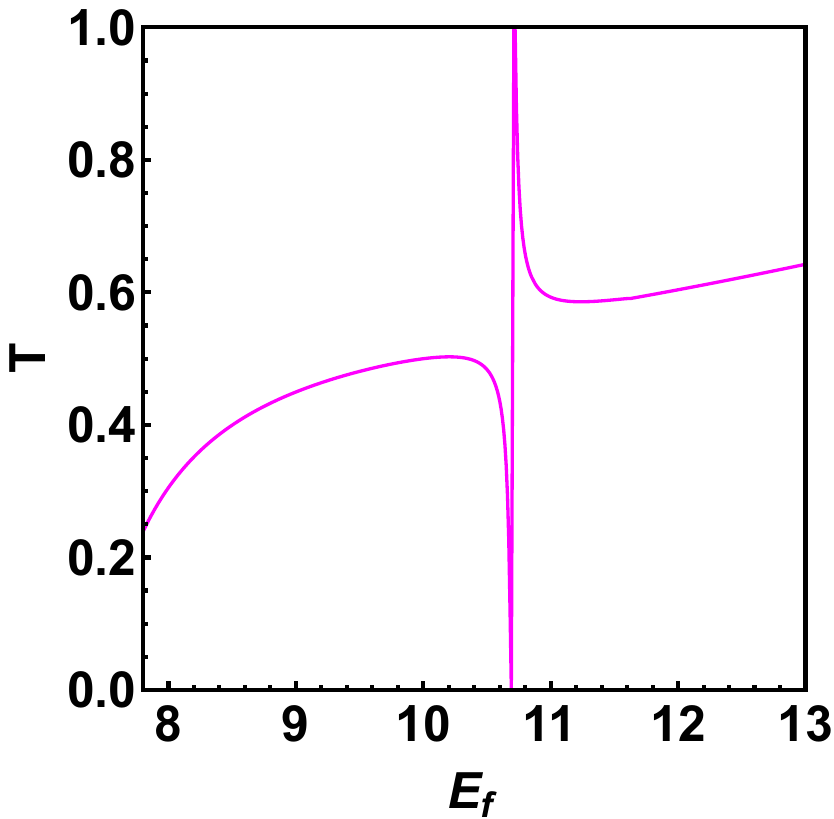}}
\caption{\label{3d_tf_v0}
3d QBT: Panels (a), (b), (c), and (d) show the total Floquet transmission coefficient $T$ as a function of $E_f$ (in units of meV), for $ V_0=5, \,10,\,15,\,20$ meV, respectively.
The remaining parameter values are kept fixed at $k_{y}=k_{z}=0.001\, {\text{\AA}}^{-1}$,
$\hbar\,\omega=4 $ meV, $L=3000$ {\AA}, and $V_{1}=1$ meV.}
	\end{center}	
\end{figure}

Now, let us discuss the transport features of the 3d QBT, and also discuss their similarities / differences with the 2d case.
The numerical results for the transmission coefficient $T$, the shot noise, and the derivative of the shot noise, as functions of the $E_f$, are shown in Figs.~\ref{large3d} and \ref{large3d2}.
Similar to the type $1$ FRPs of the 2d case, the $E_f$ value of the FRP increases with $k_{y}$ for a fixed value of $k_{z}$. Due to the symmetry between the $k_y$ and $k_z$ momenta, the same behaviour is expected for increasing value of the momentum along any direction in the $yz$-plane.
We have performed our calculations over a large interval of momenta, namely varying
$k_y,\,k_z $ in the interval between $0.0002 \,{\text{\AA}}^{-1}$ and $0.0015 \,{\text{\AA}}^{-1}$.
But, we did not find the type $2$ FRPs in this 3d case.
The shot noise patterns also differ from that in the 2d case, as seen in Figs.~\ref{noise3d_1}, \ref{derinoise3d_1}, \ref{noise3d_2}, and \ref{derinoise3d_2}. 

Fig.~\ref{3d_tf_v0} shows how the Fano resonance depends on the static potential $V_0$. The evolution of the FRP (which is only of type $1$) is similar to what is seen for the 2d QBT.

Again, the reasons for the differences in behaviour of the 3d QBT FRPs from those in 2d QBT as well as other mesoscopic Hamiltonians will be discussed in the next subsection, where the role of the bound state spectra will be discussed.

\subsection{Identifying Fano resonances with (quasi)bound states}
\label{secb}

\begin{figure}[h]
{\includegraphics[width=0.4 \textwidth]{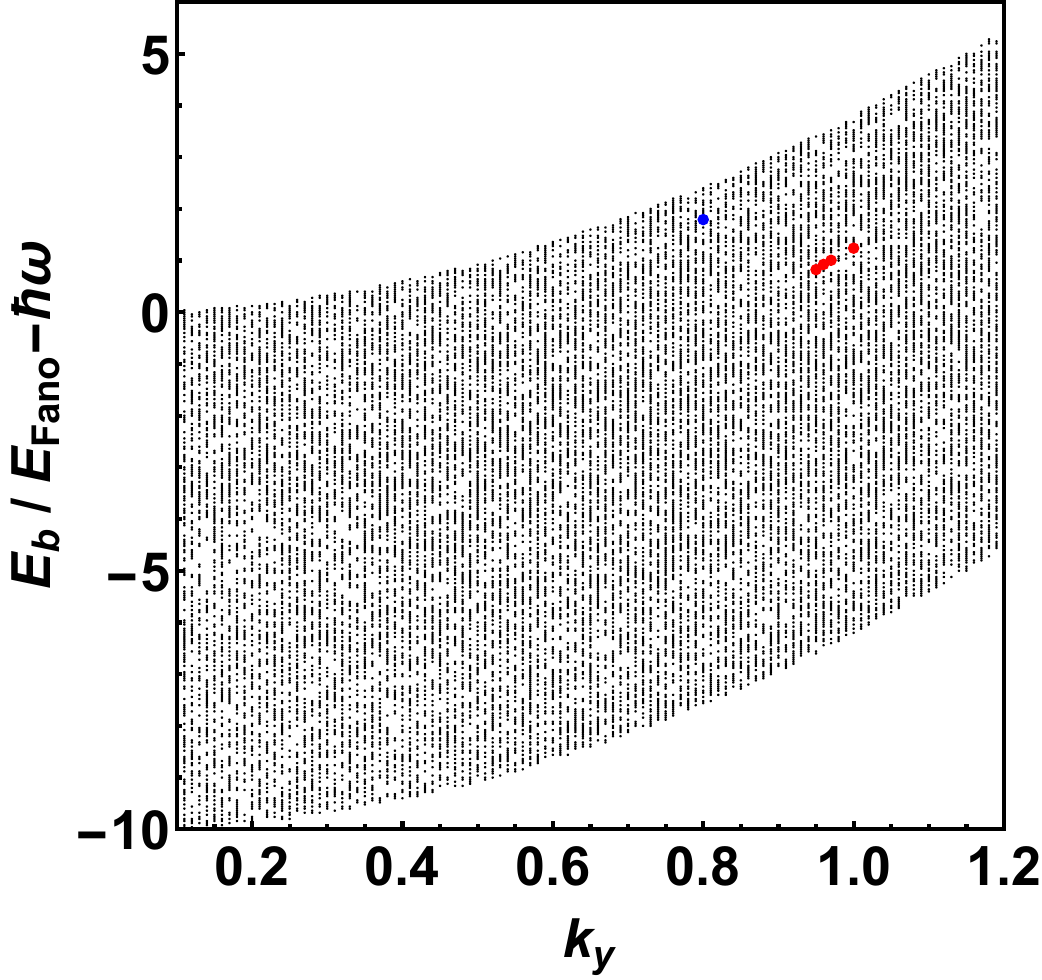}}	
\caption{2d QBT: The black dots represent the energies of the bound states $E_{b}$ (in meV) of the static quantum well, as we vary $k_y$ (in units of $10^{-3} \,{\text{\AA}} ^{-1}$). The red and blue dots represent the values of  $E_{\text{Fano}}- \hbar\,\omega$ for some of the type $1$ and type $2$ FRPs, respectively, shown in the earlier plots for 2d QBT.\label{2dquasi}}
\end{figure}
\begin{figure}[h]	
\subfloat[\label{3d_quasi}]{\includegraphics[width=0.45 \textwidth]{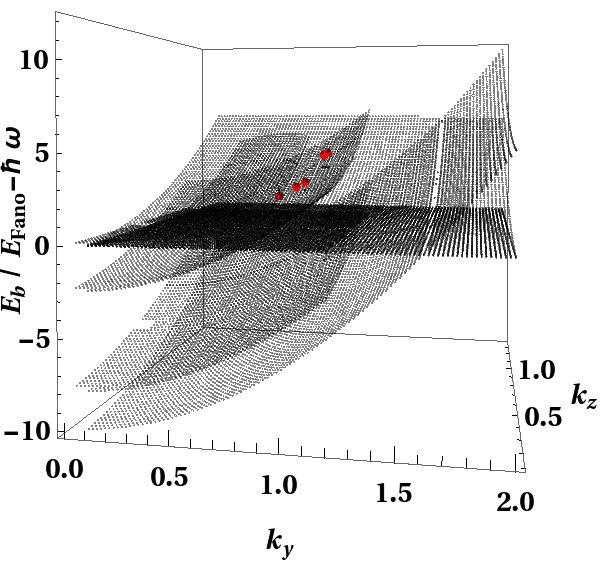}}	\\
\subfloat[\label{3d_quasi2}]{\includegraphics[width=0.3 \textwidth]{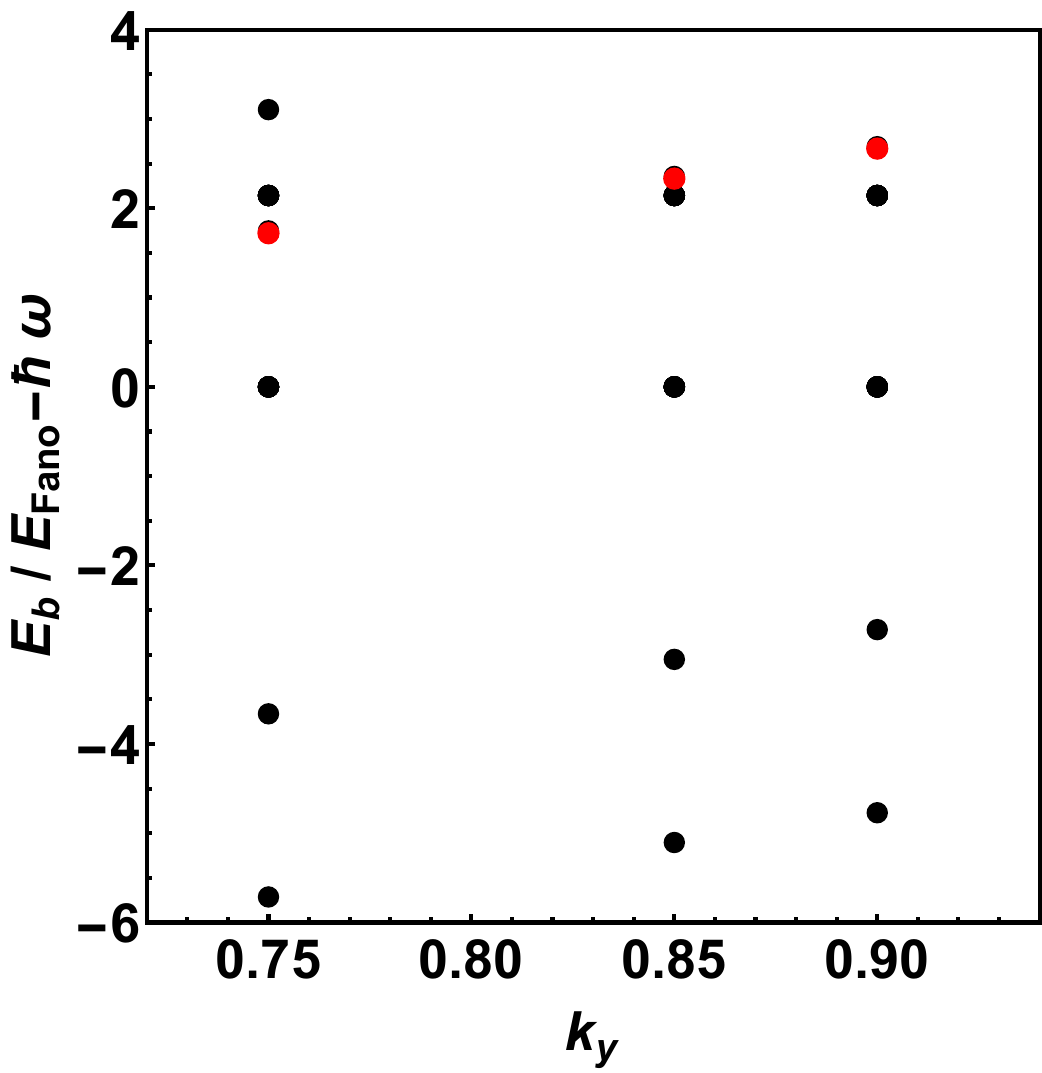}}
\hspace{3 cm}
\subfloat[\label{3d_quasi3}]{\includegraphics[width=0.3 \textwidth]{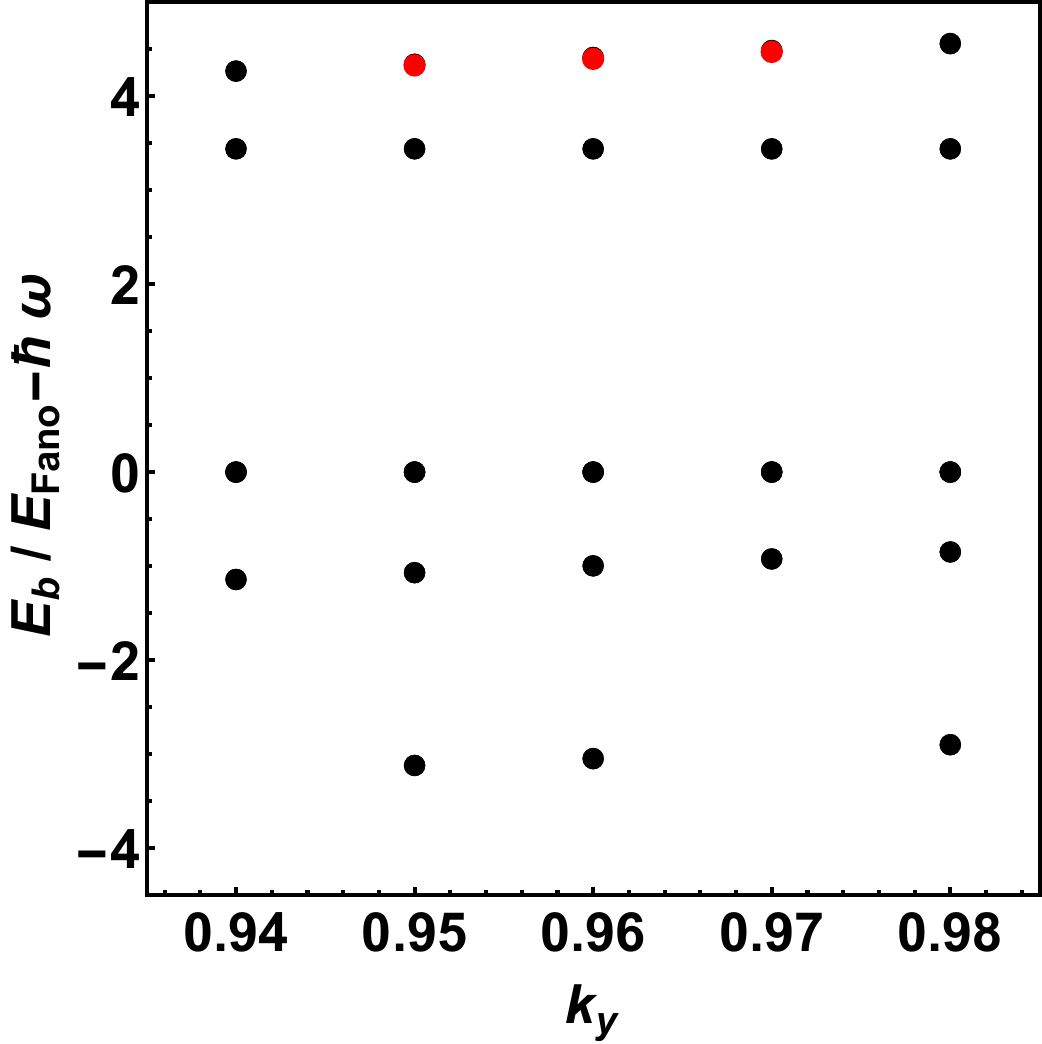}}
 \caption{\label{figquasi}
3d QBT: The black dots represent the energies of the bound states $E_{b}$ (in meV) of the static quantum well, as we vary both $k_y$ and $k_z$ (in units of $10^{-3} \,{\text{\AA}} ^{-1}$). The red dots represent the values of  $E_{\text{Fano}}- \hbar\,\omega$ obtained from the FRPs in Figs.~\ref{large3d} and \ref{large3d2}. Panels (b) and (c) show the bound state energies (in mev) at $k_{z}=0.00075 \,{\text{\AA}} ^{-1}$ and $k_{z}=0.00095 \,{\text{\AA}} ^{-1}$, respectively, as we vary $k_y$ (in units of $10^{-3} \,{\text{\AA}} ^{-1}$). In all the panels, the numerical parameters used for the driven well are: $\hbar\,\omega=4 $ meV, $L=3000$ {\AA}, $V_0=10$ meV, and $V_{1}=1$ meV.}
\end{figure}


We first consider the bound states within the static quantum well for the 2d QBT.
The $n=m=0$ modes in Eqs.~\eqref{eqwv1} and \eqref{eqwv2} represent this scenario.
The roots of $E_0$ obtained from the secular equation
\begin{align}
\label{eq2dbound}
& \begin{vmatrix}
 \gamma _{1,0} \,e^{ -\frac{k_{0} L}{2}} & 
-\gamma _{2,0} \,e^{ -\frac{i q_0 L}{2}} & -\gamma _{2,0} \,e^{ \frac{i q_0 L}{2}} & 0 \\
0 & \gamma _{2,0} \,e^{ \frac{i q_0 L}{2}} & 
\gamma _{2,0} \,e^{ -\frac{i q_0 L}{2}} & 
-\gamma _{1,0}\, e^{ -\frac{k_{0} L}{2}} \\
\frac{\gamma _{1,0} \,e^{ -\frac{k_{0} L}{2}} k_{y}} {i \,k_{0}} & 
\frac{\gamma_{2,0}\, e^{ -\frac{i q_0 L}{2}} k_{y}}{q_0} & 
-\frac{\gamma _{2,0}\, e^{ \frac{i q_0 L}{2}} k_{y}}{q_0} & 0 \\
0 & -\frac{\gamma _{2,0} \,e^{ \frac{i q_0 L}{2}} k_{y}}{q_0} & 
\frac{\gamma _{2,0} \,e^{ -\frac{i q_0 L}{2}} k_{y}}{q_0} & \frac{\gamma _{1,0}\, 
e^{ -\frac{k_{0} L}{2}} k_{y}} {i\, k_{0}} \\
\end{vmatrix}=0 \nn
& \Rightarrow
\tan \left( L\, \sqrt{\frac{2 \,\mu \, (E_b + V_0)}
{{\hbar}^2}-k_y^2} \right)
=  
\frac{\hbar^2 \sqrt{k_y^2-\frac{2 \,\mu\, E_b }{\hbar^2}} \,
\sqrt{\frac{2 \,\mu \, (E_b+V_0)}{\hbar^2}-k_y^2}}
{2 \,\mu\, E_b  +\mu \, V_0-\hbar^2 k_y^2}
\end{align}
give the energies $E_b $ of the bound states. We note that
$\frac{\hbar^2 k_{y}^2}{2\,\mu} -V_{0} \leq E_b \leq\frac{\hbar^2 k_{y}^2}{2\,\mu}$.
The black points in Fig.~\ref{2dquasi} show the numerically obtained values of the secular equation. 

In our plots for 2d QBT, an FRP is observed when the first order Floquet sideband coincides with the shallowest bound state. Hence, the locations of the FRPs are determined
from the relation $E_{Fano}-\hbar\,\omega=E_{b}$, which is equivalent to $k_{x,-1}=k_{bx}$, with $k_{bx}$ being the $x$-component of the wavevector of the quasibound level within the well. For example, in Fig.~\ref{derinoise}, the blue differential shot noise curve for $k_{y}=0.00095 \text{\AA}^{-1}$ has a sharp dip at $E_{Fano} =   4.844 $ meV, which implies that $E_{b}$ is $0.844$ meV  (with $\hbar\,\omega=4$ meV). The numerical analysis of Eq.~\eqref{eq2dbound} indeed gives a bound state at $E_{b}=0.844$ meV for the static well. This explains the Fano resonance at $E_f= 4.844 $ meV. Fig.~\ref{2dquasi} shows that there are many static bound states for a given $k_{y}$. 
If we increase the driving frequency $\omega$, the sideband energy interval $\hbar\,\omega$ will increase, and bound states with higher values of $E_b$ (i.e. deeper bound states) will then be activated to produce Fano resonances in the transmission spectrum.

Next, let us consider the bound states within the static quantum well for the 3d QBT.
The $n=m=0$ modes in Eq.~\eqref{eq3dwave} represent this scenario.
The roots of $E_0$ obtained from the secular equation
\begin{align}
\label{eq3dbound}
\begin{vmatrix}
  f_{i011} \,e^{-\frac{k_0 L}{2}} &f_{i021}  \,e^{-\frac{k_0 L}{2}} & -\tilde f_{i011}  \,e^{\frac{-i q_0 L}{2}} & -\tilde f_{i021}  \,e^{\frac{-i q_0 L}{2}} & -\tilde f_{o011}  \,e^{\frac{-i q_0 L}{2}} & -\tilde f_{o021}  \,e^{\frac{-i q_0 L}{2}} & 0 & 0 \\
  f_{i012}  \,e^{-\frac{k_0 L}{2}} & f_{i022}  \,e^{-\frac{k_0 L}{2}} & -\tilde f_{i012}  \,e^{\frac{-i q_0 L}{2}} & -\tilde f_{i022}  \,e^{\frac{-i q_0 L}{2}} & -\tilde f_{o012}  \,e^{\frac{-i q_0 L}{2}} & -\tilde f_{o022}  \,e^{\frac{-i q_0 L}{2}} & 0 & 0 \\
  f_{i013}  \,e^{-\frac{k_0 L}{2}} & f_{i023}  \,e^{-\frac{k_0 L}{2}} & -\tilde f_{i013}  \,e^{\frac{-i q_0 L}{2}} & -\tilde f_{i023}  \,e^{\frac{-i q_0 L}{2}} & -\tilde f_{o013}  \,e^{\frac{-i q_0 L}{2}} & -\tilde f_{o023}  \,e^{\frac{-i q_0 L}{2}} & 0 & 0 \\
  f_{i014}  \,e^{-\frac{k_0 L}{2}} & f_{i024}  \,e^{-\frac{k_0 L}{2}} & -\tilde f_{i014}  \,e^{\frac{-i q_0 L}{2}} & -\tilde f_{i024}  \,e^{\frac{-i q_0 L}{2}} & -\tilde f_{o014}  \,e^{\frac{-i q_0 L}{2}} & -\tilde f_{o024}  \,e^{\frac{-i q_0 L}{2}} & 0 & 0 \\
  0 & 0 & \tilde f_{i011}  \,e^{\frac{-i q_0 L}{2}} & \tilde f_{i021}  \,e^{\frac{-i q_0 L}{2}} & \tilde f_{o011}  \,e^{\frac{-i q_0 L}{2}} & \tilde f_{o021}  \,e^{\frac{-i q_0 L}{2}} & -f_{o011}  \,e^{-\frac{k_0 L}{2}} & -f_{o021}  \,e^{-\frac{k_0 L}{2}} \\
  0 & 0 & \tilde f_{i012}  \,e^{\frac{-i q_0 L}{2}} & \tilde f_{i022}  \,e^{\frac{-i q_0 L}{2}} & \tilde f_{o012}  \,e^{\frac{-i q_0 L}{2}} & \tilde f_{o022}  \,e^{\frac{-i q_0 L}{2}} & -f_{o012}  \,e^{-\frac{k_0 L}{2}} & -f_{o022}  \,e^{-\frac{k_0 L}{2}} \\
  0 & 0 & \tilde f_{i013}  \,e^{\frac{-i q_0 L}{2}} & \tilde f_{i023}  \,e^{\frac{-i q_0 L}{2}} & \tilde f_{o013}  \,e^{\frac{-i q_0 L}{2}} & \tilde f_{o023}  \,e^{\frac{-i q_0 L}{2}} & -f_{o013}  \,e^{-\frac{k_0 L}{2}} & -f_{o023}  \,e^{-\frac{k_0 L}{2}} \\
  0 & 0 & \tilde f_{i014}  \,e^{\frac{-i q_0 L}{2}} & \tilde f_{i024}  \,e^{\frac{-i q_0 L}{2}} & \tilde f_{o014}  \,e^{\frac{-i q_0 L}{2}} & \tilde f_{o024}  \,e^{\frac{-i q_0 L}{2}} & -f_{o014}  \,e^{-\frac{k_0 L}{2}} & -f_{o024}  \,e^{-\frac{k_0 L}{2}} \\
  \end{vmatrix}=0
  \end{align}
give the energies $E_b $ of the bound states.
The black points in Figs.~\ref{3d_quasi}, \ref{3d_quasi2}, and \ref{3d_quasi3} show the numerically obtained values of the secular equation. The red dots represent the values of  $E_{\text{Fano}}- \hbar\,\omega$ obtained from the FRPs in Figs.~\ref{large3d} and \ref{large3d2}. We have a high density of points for the 3d case (due to the extra dimension $k_{z}$) in comparison with 2d case. This certainly affects the  lower / upper boundary for the appearance of the type $1$ (type $2$) FRPs in the momentum space. Our simulations suggest that the upper boundary for obtaining the type $2$ FRPs might be very low, as we did not find these in the interval $0.0002\,\text{\AA}^{-1}-0.0015\,\text{\AA}^{-1}$. It might be possible that the type $2$ FRPs exist at very low momenta.

From the above discussions, it is clear that the behaviour of the FRPs is primarily controlled by the bound state spectra of the corresponding systems. These, in turn, are determined by the structure of the Hamiltonian. Different Hamiltonians with different dispersions (e.g. linear versus quadratic), dimensionality (2d versus 3d), and different spinorial structures of the eigenfunctions, give rise to distinct transcendental equations determining the bound states.
Clearly, the transcendental equations in Eqs.~\eqref{eq2dbound} and \eqref{eq3dbound} are different from each other, as well from those in Ref.~\cite{Zhu15} (where the cases of 2d electron gas and graphene are discussed). Both the matrix structure of Hamiltonian, and the powers of momenta appearing in the dispersion, feed into the solutions.

 \section{Summary and outlook}
\label{secsum} 
In this paper, we have devised a formalism to chart out the Fano resonance features in the transmission spectra of the electrons passing through a periodically driven quantum well in 2d and 3d QBT semimetals. We have looked at the nonadiabatic limit where the Floquet scattering framework can be used. From our detailed analysis, we have identified the FRPs with the bound states of the systems. The Fano resonance spectra show the presence of various kinds of FRPs, including those having narrow bandwidths. Extremely narrow bandwidth FRPs can have useful applications for designing advanced dynamic reconfigurable devices \cite{Karmakar19}. We have also compared our results with systems involving free electron gas, graphene, and pseudospin-1 Dirac-Weyl semimetals.

In experiments, the Fano resonances can be captured through the measurement of the pumped shot noise (see, for example Ref.~\cite{shot_noise}), which we have have computed in our paper. Although the shot noise for the static case has been experimentally studied for HgTe quantum wells \cite{shot_hg}, we are not aware of specific experiments involving harmonic drives carried out with such materials. We hope our theoretical investigations will stimulate such experimental work on these very interesting systems.

In future, it will be useful to look at the Floquet scattering properties in the presence of disorder \cite{rahul-sid,*ipsita-rahul,*ips-qbt-sc} and/or magnetic fields \cite{mansoor,ips3by2,ipsita-aritra}. Another direction is to examine the effects of anisotropy as well as particle-hole symmetry-breaking terms in the Hamiltonian.
It will also be interesting to investigate the effect of periodic potential in presence of interactions, which can have drastic effects like destroying quantization of various physical quantities in the topological phases \cite{PhysRevResearch.2.013069,kozii,Mandal_2020}, or the emergence of strongly correlated phases~\cite{ips-seb,MoonXuKimBalents,rahul-sid,ipsita-rahul,ips-qbt-sc} where quasiparticle description of transport breaks down \cite{ips-subir,ipsc2,ips-mem-mat}.

 \section*{Acknowledgments} 
 
We thank Tanay Nag for useful discussions and for participating in the initial stages of the project. We are also grateful to Rui Zhu for explaining the shot noise formalism for graphene. 

\bibliography{ref}

\appendix

\section{Floquet scattering matrix for the 2d QBT case}
\label{cal_2d}

The continuity of the components of the wavefunction, as given in Eqs.~\eqref{eqwv1} and \eqref{eqwv2}, at the boundaries $ x =\pm L/2$,
enables us to derive the values of the unknown coefficients $ A_{n}^{i},\, A_{n}^{o},\,  B_{n}^{i},\,  B_{n}^{o}, \, a_m,$ and $ b_m$. 
For simplifying the equations, we define some matrices whose components are given by:  
\begin{align}
\label{30}
& \left({   M_{1s}}^{\pm} \right )_{nm}
=   \frac{\gamma_{2}}{\gamma_{1}} \left[ \left (1+\frac{k_n}{q_{m}} \right ) \,e^{-\frac{i q_{m}L} {2}} \pm    
\left (1-\frac{k_n}{q_{m}} \right ) \,e^{ \frac{iq_{m}L} {2}} \right]
 J_{n-m}\left( \frac{V_{1}}{\hbar \omega}\right) ,
 \quad M_{i}^{nm}= e^{-ik_nL}\, \delta_{n,m}\,,
\nn & 
\left({   M_{2s}}^{\pm} \right)_{nm}
= \frac{\gamma_{3}}{\gamma_{1}}
\left[  \left(1-\frac{k_n\, q_{m}}{k_{y}^2} \right) \,e^{-\frac{iq_{m}L}{2}} \pm     
\left (1+\frac{k_n\, q_{m}}{k_{y}^2} \right ) \,e^{\frac{iq_{m}L} {2} } \right] 
J_{n-m}\left( \frac{V_{1}}{\hbar \omega}\right) ,
\quad { C}_{m}^{\pm}= a_{m} \pm b_{m}\,,
\nn
& {   M_{r}} ^{nm}= 2\,e^{-\frac{ik_nL } {2}} \delta_{n,m}\,,\quad 
\left({   M_{1c}}^{\pm} \right)_{nm}= 
\frac{\gamma_{2}}{\gamma_{1}}\,
e^{-\frac { i(k_n \pm q_{m})L} {2}}J_{n-m}\left( \frac{V_{1}}{\hbar \omega}\right) ,
\quad
\left({   M_{2c}}^{\pm} \right)_{nm}
= \frac{\gamma_{3}}{\gamma_{1}}\,
e^{- \frac{i \,(k_n \pm q_{m})L  } {2} }
J_{n-m}\left( \frac{V_{1}}{\hbar \omega}\right) .
\end{align} 
From the continuity conditions, we then get:
\begin{align}\label{35}
{   M_{r}}\cdot({   A^{i}} \pm  {   B^{i}}) 
&=  {  M_{1s}}^{\pm} \cdot {   C}^{\pm} \, \Theta(E_m+ V_0) 
+{  M_{2s}}^{\pm} \cdot {   C}^{\pm}\, \Theta(-V_0-E_m) \,,\nn      
{  A}^{o} &=
\left [  { M}_{jc}^{+} 
\cdot{
\left \lbrace \left ( {M}_{js}^{+} \right )^{-1}
+ \left ( {   M}_{js}^{-} \right )^{-1}\right \rbrace }
+{   M}_{jc}^{-} \cdot{\left \lbrace  ( {   M}_{js}^{+})^{-1}
-\left ( {   M}_{js}^{-} \right )^{-1}\right \rbrace } 
\right ]  
\frac{{   M}_{r}  \cdot   {   A}^{i}}{2} 
 -{   M_{i}}\cdot{   A^{i}} \nn
  & \quad  +\left [  {   M}_{jc}^{+} \cdot
  {\left \lbrace  \left( {   M}_{js}^{+} \right )^{-1}
  -\left ( {   M}_{js}^{-} \right )^{-1}\right \rbrace  }
 +{   M}_{jc}^{-} \cdot{\left \lbrace  \left( {   M}_{js}^{+} \right )^{-1}
 + \left( {   M}_{js}^{-} \right )^{-1}\right \rbrace } \right ]
   \frac{{   M}_{r}  \cdot   {   B}^{i}}{2}  \nn 
& \equiv  {   A^{i}} \cdot {   M}_{AA}
+ {   B^{i}} \cdot  {   M}_{AB}\,,\nn
{  B}^{o} &=\left [  { M}_{jc}^{-} \cdot
\left \lbrace ( {M}_{js}^{+})^{-1}+( {M}_{js}^{-})^{-1}\right \rbrace
+{ M}_{jc}^{+} \cdot \left \lbrace ( {   M}_{js}^{+})^{-1}-( {   M}_{js}^{-})^{-1}
\right \rbrace  \right ]
  \frac{{  M}_{r}  \cdot   { A}^{i}}{2}  \\
\nonumber & 
\quad + \left [  { M}_{jc}^{+} \cdot \left \lbrace ( {  M}_{js}^{+})^{-1}+( { M}_{js}^{-})^{-1}\right \rbrace 
+{M}_{jc}^{-} \cdot \left \lbrace ( {   M}_{js}^{+})^{-1}-( {   M}_{js}^{-})^{-1}\right \rbrace \right ]  \frac{{M}_{r}  \cdot   {B}^{i}}{2}-{ M_{i}}\cdot   B^{i}  \nn
& \equiv {   A^{i}} \cdot  {   M}_{BA}
+{   B^{i}} \cdot {   M}_{BB}\,,
\end{align}
where $j=1,2$. We have to set $j=1$ for $E_m>-V_0$ and $j=2$ for $E_m<-V_0$.
This finally gives us the matrix equation:
\begin{align}
\begin{pmatrix} 
{  A}^{o}\\{  B}^{o}
\end{pmatrix}   =  \begin{pmatrix} 
{  M}_{AA} & {  M}_{AB} \\ {  M}_{BA} & {  M}_{BB}
\end{pmatrix} 
\begin{pmatrix} 
{  A}^{i}\\{  B}^{i}
\end{pmatrix} \,.
\end{align}
For the $n^{\text{th}}$ Floquet band, the above matrix equation reduces to: 
\begin{align}
\begin{pmatrix} 
{  A}_{n}^{o}\\{  B}_{n}^{o}
\end{pmatrix}   = \sum \limits_{m=-\infty}^\infty \mathcal{S}_{nm}  
  \begin{pmatrix} 
  A_{m}^{i}\\{  B}_{m}^{i}
 \end{pmatrix}\,,
\end{align}	  
where we can determine $\mathcal{S}_{nm}$ from ${M}_{AA}, \, {M}_{AB},\,{M}_{BB},$ and ${M}_{BA}$.


\section{Floquet scattering matrix for the 3d QBT case} 
\label{cal_3d}

 The continuity of the components of the wavefunction, as given in Eq.~\eqref{eq3dwave}, at the boundaries $ x =\pm L/2$,
enables us to derive the values of the unknown coefficients $ A_{n,1}^{i},\,A_{n,2}^{i},\, A_{n,1}^{o},\,A_{n,2}^{o},$ $ B_{n,1}^{i}
,\,B_{n,2}^{i},\,  B_{n,1}^{o}, \, B_{n,2}^{o},$ $\alpha_{m,1},\, \alpha_{m,2},$
$\beta_{m,1},$ and $ \beta_{m,2}$. 
  For simplifying the equations, we define some matrices whose components are given by:  
 \begin{align}
 {v}_{11}^{nm}&= \frac{ (k_{n}-q_{m})\, (q_{m}+i k_{y}) \sqrt{\chi_{n}^3 (k_{z}+\chi_{n})} 
\, e^{-\frac{ i  (k_{n}+q_{m}) L}{2}} 
 \left[ k_{n} (k_{z}+\chi_{m})+q_{m} (k_{z}-\chi_{n})+i k_{y} (\chi_{n}+\chi_{m}) \right ]
J_{n-m}\left( \frac{V_{1}}{\hbar \omega}\right) }
 {4 \,k_{n}\, \chi_{n} \,(k_{n}-i k_{y}) \,(q_{m}-i k_{y}) \sqrt{\chi_{m}^3 (k_{z}+\chi_{m})}} \,,\nn
{v}_{12}^{nm}
&= \frac{ (k_{n}-q_{m})\, (q_{m}+i k_{y}) \sqrt{\chi_{n}^3 
(k_{z}+\chi_{n})} \,e^{-\frac{i  (k_{n}+q_{m}) L} {2}} 
\left[ k_{n} (k_{z}-\chi_{m})+q_{m} (k_{z}-\chi_{n})+i k_{y} (\chi_{n}-\chi_{m})
\right ]
J_{n-m}\left( \frac{V_{1}}{\hbar \omega}\right)}
{4 \,k_{n} \,\chi_{n} \,(k_{n}-i k_{y})\, (q_{m}-i k_{y}) \sqrt{\chi_{m}^3 (\chi_{m}-k_{z})}} \,,\nn
 {v}_{13}^{nm}&=\frac{ (k_{n}+q_{m}) \,(q_{m}-i k_{y}) 
 \sqrt{\chi_{n}^3 (k_{z}+\chi_{n})} \,e^{-\frac{ i  (k_{n} - q_{m}) L} {2}} 
 \left [ k_{n} (k_{z}+\chi_{m})+q_{m} (\chi_{n}-k_{z})+i k_{y} (\chi_{n}+\chi_{m})
 \right ]
J_{n-m}\left( \frac{V_{1}}{\hbar \omega}\right) }
 {4 \,k_{n}\, \chi_{n} \,(k_{n}-i k_{y})\, (q_{m}+i k_{y}) \sqrt{\chi_{m}^3 (k_{z}+\chi_{m})}} \,,\nn
{v}_{14}^{nm}&=\frac{ (k_{n}+q_{m})\, (q_{m}-i k_{y}) \sqrt{\chi_{n}^3 (k_{z}+\chi_{n})} 
 \, e^{-\frac{i  (k_{n} - q_{m}) L }{2}} (k_{n} (k_{z}-\chi_{m})+q_{m} (\chi_{n}-k_{z})+i k_{y} (\chi_{n}-\chi_{m})) J_{n-m}\left( \frac{V_{1}}{\hbar \omega}\right)}
 {4 \,k_{n}\, \chi_{n} \,(k_{n}-i k_{y}) \,(q_{m}+i k_{y}) \sqrt{\chi_{m}^3 (\chi_{m}-k_{z})}} \,,
\end{align} 
 \begin{align}
 {v}_{21}^{nm}&= -\frac{ (k_{n}-q_{m})\, (q_{m}+i k_{y}) \sqrt{\chi_{n}^3 (\chi_{n}-k_{z})}  e^{-\frac{i  (k_{n}+q_{m}) L }{2}} (k_{n} (k_{z}+\chi_{m})+q_{m} (k_{z}+\chi_{n})-i k_{y} (\chi_{n}-\chi_{m})) J_{n-m}\left( \frac{V_{1}}{\hbar \omega}\right)}
 {4\, k_{n}\, \chi_{n} \,(k_{n}-i k_{y})\, (q_{m}-i k_{y}) \sqrt{\chi_{m}^3 (k_{z}+\chi_{m})}}
 \,,\nn
 {v}_{22}^{nm}&= -\frac{ (k_{n}-q_{m})\, (q_{m}+i k_{y}) \sqrt{\chi_{n}^3 (\chi_{n}-k_{z})}  e^{-\frac{i  (k_{n}+q_{m}) L }{2}} 
 \left[ k_{n} (k_{z}-\chi_{m})+q_{m} (k_{z}+\chi_{n})-i k_{y} (\chi_{n}+\chi_{m}) \right ]
 J_{n-m}\left( \frac{V_{1}}{\hbar \omega}\right)}
 {4\, k_{n}\, \chi_{n} \,(k_{n}-i k_{y})\, (q_{m}-i k_{y}) \sqrt{\chi_{m}^3 (\chi_{m}-k_{z})}}\,,
 \nn
 {v}_{23}^{nm}&=\frac{ (k_{n}+q_{m})\, (q_{m}-i k_{y}) \sqrt{\chi_{n}^3 (\chi_{n}-k_{z})}  e^{-\frac{i  (k_{n} - q_{m}) L }{2}} 
 \left[ -k_{n} (k_{z}+\chi_{m})+q_{m} (k_{z}+\chi_{n})+i k_{y} (\chi_{n}-\chi_{m}) \right  ]
 J_{n-m}\left( \frac{V_{1}}{\hbar \omega}\right)}
 {4\, k_{n}\, \chi_{n} \,(k_{n}-i k_{y})\, (q_{m}+i k_{y}) \sqrt{\chi_{m}^3 (k_{z}+\chi_{m})}}\,,\nn
{v}_{24}^{nm}&=\frac{ (k_{n}+q_{m})\, (q_{m}-i k_{y}) \sqrt{\chi_{n}^3 (\chi_{n}-k_{z})}  e^{-\frac{i  (k_{n} - q_{m}) L }{2}} 
\left[ k_{n} (\chi_{m}-k_{z})+q_{m} (k_{z}+\chi_{n})+i k_{y} (\chi_{n}+\chi_{m}) \right ]
J_{n-m}\left( \frac{V_{1}}{\hbar \omega}\right)}
{4\, k_{n}\, \chi_{n} \,(k_{n}-i k_{y})\, (q_{m}+i k_{y}) \sqrt{\chi_{m}^3 (\chi_{m}-k_{z})}}\,,
 \end{align}
 \begin{align}
{v}_{31}^{nm}&= \frac{ (k_{n}+q_{m})\, (q_{m}+i k_{y}) \sqrt{\chi_{n}^3 (k_{z}+\chi_{n})} 
\, e^{-\frac{i  (k_{n} - q_{m}) L }{2}} 
\left[ k_{n} (k_{z}+\chi_{m})+q_{m} (\chi_{n}-k_{z})-i k_{y} (\chi_{n}+\chi_{m}) \right ]
J_{n-m}\left( \frac{V_{1}}{\hbar \omega}\right)}
{4\, k_{n}\, \chi_{n} \,(k_{n}+i k_{y})\, (q_{m}-i k_{y}) \sqrt{\chi_{m}^3 (k_{z}+\chi_{m})}} \,,\nn
 {v}_{32}^{nm}
 &= \frac{ (k_{n}+q_{m})\, (q_{m}+i k_{y}) \sqrt{\chi_{n}^3 (k_{z}+\chi_{n})}  
 \,e^{-\frac{i  (k_{n} - q_{m}) L }{2}} 
 \left[ k_{n} (k_{z}-\chi_{m})+q_{m} (\chi_{n}-k_{z})-i k_{y} (\chi_{n}-\chi_{m}) \right ]
J_{n-m}\left( \frac{V_{1}}{\hbar \omega}\right)}
 {4\, k_{n}\, \chi_{n} \,(k_{n}+i k_{y})\, (q_{m}-i k_{y}) \sqrt{\chi_{m}^3 (\chi_{m}-k_{z})}}\,,\nn
 {v}_{33}^{nm}&=
 \frac{ (k_{n}-q_{m})\, (q_{m}-i k_{y}) \sqrt{\chi_{n}^3 (k_{z}+\chi_{n})}  
 \,e^{-\frac{i  (k_{n}+q_{m}) L }{2}} 
 \left  [ k_{n} (k_{z}+\chi_{m})+q_{m} (k_{z}-\chi_{n})-i k_{y} (\chi_{n}+\chi_{m}) \right ]
 J_{n-m}\left( \frac{V_{1}}{\hbar \omega}\right)}
 {4\, k_{n}\, \chi_{n} \,(k_{n}+i k_{y})\, (q_{m}+i k_{y}) \sqrt{\chi_{m}^3 (k_{z}+\chi_{m})}}
\,,\nn
 {v}_{34}^{nm}
 &=-\frac{ (k_{n}-q_{m})\, (q_{m}-i k_{y}) \sqrt{\chi_{n}^3 (k_{z}+\chi_{n})}  
 \, e^{-\frac{i  (k_{n}+q_{m}) L }{2}} 
 \left [k_{n} (\chi_{m}-k_{z})+q_{m} (\chi_{n}-k_{z})+i k_{y} (\chi_{n}-\chi_{m})
 \right ]
 J_{n-m}\left( \frac{V_{1}}{\hbar \omega}\right)}
 {4\, k_{n}\, \chi_{n} \,(k_{n}+i k_{y})\, (q_{m}+i k_{y}) \sqrt{\chi_{m}^3 (\chi_{m}-k_{z})}}\,,
 \end{align}
 \begin{align}
{v}_{41}^{nm}&= \frac{ (k_{n}+q_{m})\, (q_{m}+i k_{y}) \sqrt{\chi_{n}^3 
 (\chi_{n}-k_{z})}  \,e^{-\frac{i  (k_{n} - q_{m}) L }{2}} 
\left[ -k_{n} (k_{z}+\chi_{m})+q_{m} (k_{z}+\chi_{n})-i k_{y} (\chi_{n}-\chi_{m}) \right ]
 J_{n-m}\left( \frac{V_{1}}{\hbar \omega}\right)}
 {4\, k_{n}\, \chi_{n} \,(k_{n}+i k_{y})\, (q_{m}-i k_{y}) \sqrt{\chi_{m}^3 (k_{z}+\chi_{m})}}
 \,,\nn
  {v}_{42}^{nm}&= \frac{ (k_{n}+q_{m})\, (q_{m}+i k_{y}) \sqrt{\chi_{n}^3 (\chi_{n}-k_{z})}  
  \,e^{-\frac{i  (k_{n} - q_{m}) L }{2}} \left[ k_{n} (\chi_{m}-k_{z})+q_{m} (k_{z}+\chi_{n})-i k_{y} (\chi_{n}+\chi_{m}) \right ]
J_{n-m}\left( \frac{V_{1}}{\hbar \omega}\right)  }
  {4\, k_{n}\, \chi_{n} \,(k_{n}+i k_{y})\, (q_{m}-i k_{y}) \sqrt{\chi_{m}^3 (\chi_{m}-k_{z})}}
\,,\nn
{v}_{43}^{nm}&=-\frac{ (k_{n}-q_{m})\, (q_{m}-i k_{y}) \sqrt{\chi_{n}^3 (\chi_{n}-k_{z})}  
\,e^{-\frac{i  (k_{n}+q_{m}) L }{2}} \left[ k_{n} (k_{z}+\chi_{m})+q_{m} (k_{z}+\chi_{n})+i k_{y} (\chi_{n}-\chi_{m})\right ]
J_{n-m}\left( \frac{V_{1}}{\hbar \omega}\right)}
{4\, k_{n}\, \chi_{n} \,(k_{n}+i k_{y})\, (q_{m}+i k_{y}) \sqrt{\chi_{m}^3 (k_{z}+\chi_{m})}}
\,,\nn
 {v}_{44}^{nm}&=
 -\frac{ (k_{n}-q_{m})\, (q_{m}-i k_{y}) \sqrt{\chi_{n}^3 (\chi_{n}-k_{z})} \,
  e^{-\frac{i  (k_{n}+q_{m}) L }{2}} \left[ k_{n} (k_{z}-\chi_{m})+q_{m} (k_{z}+\chi_{n})+i k_{y} (\chi_{n}+\chi_{m})\right ] 
 J_{n-m}\left( \frac{V_{1}}{\hbar \omega}\right)  }
  {4\, k_{n}\, \chi_{n} \,(k_{n}+i k_{y})\, (q_{m}+i k_{y}) \sqrt{\chi_{m}^3 (\chi_{m}-k_{z})}}\,,
 \end{align}
 \begin{align}\label{107}
{u}_{11}^{nm} & =\frac{ (k_{n}+q_{m})\, (q_{m}+i k_{y})\,
 e^{-\frac{i q_{m} L}{2}} \sqrt{\chi_{n}^3 (k_{z}+\chi_{n})} \left [ k_{n} (k_{z}+\chi_{m})+q_{m} (\chi_{n}-k_{z})-i k_{y} (\chi_{n}+\chi_{m}) \right ] J_{n-m}\left( \frac{V_{1}}{\hbar \omega}\right)}
 {4\, k_{n}\, \chi_{n} \,(k_{n}+i k_{y})\, (q_{m}-i k_{y}) \sqrt{\chi_{m}^3 (k_{z}+\chi_{m})}} \,,
 \nn
{u}_{12}^{nm}& = \frac{ (k_{n}+q_{m})\, (q_{m}+i k_{y}) 
 \,e^{-\frac{i q_{m} L}{2}} \sqrt{\chi_{n}^3 (k_{z}+\chi_{n})} \left[ k_{n} (k_{z}-\chi_{m})+q_{m} (\chi_{n}-k_{z})-i k_{y} (\chi_{n}-\chi_{m})
\right ] J_{n-m}\left( \frac{V_{1}}{\hbar \omega}\right)}
 {4\, k_{n}\, \chi_{n} \,(k_{n}+i k_{y})\, (q_{m}-i k_{y}) \sqrt{\chi_{m}^3 (\chi_{m}-k_{z})}}\,,\nn
 {u}_{13}^{nm}& = 
 \frac{ (q_{m}-k_{n})\, (q_{m}-i k_{y}) 
 \,e^{\frac{i q_{m} L}{2}} \sqrt{\chi_{n}^3 (k_{z}+\chi_{n})} \left [
 -k_{n} (k_{z}+\chi_{m})+q_{m} (\chi_{n}-k_{z})+i k_{y} (\chi_{n}+\chi_{m})
 \right ] J_{n-m}\left( \frac{V_{1}}{\hbar \omega}\right)}
 {4\, k_{n}\, \chi_{n} \,(k_{n}+i k_{y})\, (q_{m}+i k_{y}) \sqrt{\chi_{m}^3 (k_{z}+\chi_{m})}}
 \,,\nn
{u}_{14}^{nm}& = \frac{ (q_{m}-k_{n})\, (q_{m}-i k_{y}) 
\,e^{\frac{i q_{m} L}{2}} \sqrt{\chi_{n}^3 (k_{z}+\chi_{n})} 
\left [k_{n} (\chi_{m}-k_{z})+q_{m} (\chi_{n}-k_{z})+i k_{y} (\chi_{n}-\chi_{m})
\right ] J_{n-m}\left( \frac{V_{1}}{\hbar \omega}\right)}
{4\, k_{n}\, \chi_{n} \,(k_{n}+i k_{y})\, (q_{m}+i k_{y}) \sqrt{\chi_{m}^3 (\chi_{m}-k_{z})}}\,,
 \end{align}
 \begin{align}
{u}_{21}^{nm}& =  \frac{ (k_{n}+q_{m})\, (q_{m}+i k_{y}) \,e^{-\frac{i q_{m} L}{2}} \sqrt{\chi_{n}^3 (\chi_{n}-k_{z})} \left [ -k_{n} (k_{z}+\chi_{m})+q_{m} (k_{z}+\chi_{n})-i k_{y} (\chi_{n}-\chi_{m})
\right ] J_{n-m}\left( \frac{V_{1}}{\hbar \omega}\right)}
{4\, k_{n}\, \chi_{n} \,(k_{n}+i k_{y})\, (q_{m}-i k_{y}) \sqrt{\chi_{m}^3 (k_{z}+\chi_{m})}}\,,\nn
 {u}_{22}^{nm}= & \frac{ (k_{n}+q_{m})\, (q_{m}+i k_{y}) \,e^{-\frac{i q_{m} L}{2}} \sqrt{\chi_{n}^3 (\chi_{n}-k_{z})} \left[ 
 k_{n} (\chi_{m}-k_{z})+q_{m} (k_{z}+\chi_{n})-i k_{y} (\chi_{n}+\chi_{m})
 \right ] J_{n-m}\left( \frac{V_{1}}{\hbar \omega}\right)}
 {4\, k_{n}\, \chi_{n} \,(k_{n}+i k_{y})\, (q_{m}-i k_{y}) 
 \sqrt{\chi_{m}^3 (\chi_{m}-k_{z})}}\,,\nn
 {u}_{23}^{nm}= &\frac{ (q_{m}-k_{n})\, (q_{m}-i k_{y}) 
 \,e^{\frac{i q_{m} L}{2}} \sqrt{\chi_{n}^3 (\chi_{n}-k_{z})} \left[k_{n} (k_{z}+\chi_{m})+q_{m} (k_{z}+\chi_{n})+i k_{y} (\chi_{n}-\chi_{m})
 \right ] J_{n-m}\left( \frac{V_{1}}{\hbar \omega}\right)}
 {4\, k_{n}\, \chi_{n} \,(k_{n}+i k_{y})\, (q_{m}+i k_{y}) \sqrt{\chi_{m}^3 (k_{z}+\chi_{m})}}\,,\nn
 {u}_{24}^{nm}& = \frac{ (q_{m}-k_{n}) \,
 (q_{m}-i k_{y}) \,e^{\frac{i q_{m} L}{2}} \sqrt{\chi_{n}^3 (\chi_{n}-k_{z})} (k_{n} (k_{z}-\chi_{m})+q_{m} \left[ k_{z}+\chi_{n})+i k_{y} (\chi_{n}+\chi_{m})
 \right ] J_{n-m}\left( \frac{V_{1}}{\hbar \omega}\right)}
 {4\, k_{n}\, \chi_{n} \,(k_{n}+i k_{y})\, (q_{m}+i k_{y}) \sqrt{\chi_{m}^3 (\chi_{m}-k_{z})}}\,,
 \end{align}
 \begin{align}
 {u}_{31}^{nm} & = \frac{ (q_{m}-k_{n})\, (q_{m}+i k_{y}) \,e^{\frac{i q_{m} L}{2}} \sqrt{\chi_{n}^3 (k_{z}+\chi_{n})} \left[ -k_{n} (k_{z}+\chi_{m})+q_{m} (\chi_{n}-k_{z})-i k_{y} (\chi_{n}+\chi_{m})
 \right ] J_{n-m}\left( \frac{V_{1}}{\hbar \omega}\right)}
{4\, k_{n}\, \chi_{n} \,(k_{n}-i k_{y})\, (q_{m}-i k_{y}) \sqrt{\chi_{m}^3 (k_{z}+\chi_{m})}}\,,\nn
 {u}_{32}^{nm}& = \frac{ (q_{m}-k_{n})\, (q_{m}+i k_{y}) \,e^{\frac{i q_{m} L}{2}} \sqrt{\chi_{n}^3 (k_{z}+\chi_{n})} \left[ k_{n} (\chi_{m}-k_{z})+q_{m} (\chi_{n}-k_{z})-i k_{y} (\chi_{n}-\chi_{m})
 \right ] J_{n-m}\left( \frac{V_{1}}{\hbar \omega}\right)}
 {4\, k_{n}\, \chi_{n} \,(k_{n}-i k_{y})\, (q_{m}-i k_{y}) \sqrt{\chi_{m}^3 (\chi_{m}-k_{z})}}\,,\nn
 {u}_{33}^{nm}& = \frac{ (k_{n}+q_{m})\, (q_{m}-i k_{y}) \,e^{-\frac{i q_{m} L}{2}}
  \sqrt{\chi_{n}^3 (k_{z}+\chi_{n})} \left [ k_{n} (k_{z}+\chi_{m})+q_{m} (\chi_{n}-k_{z})+i k_{y} (\chi_{n}+\chi_{m})
 \right ] J_{n-m}\left( \frac{V_{1}}{\hbar \omega}\right)} 
  {4\, k_{n}\, \chi_{n} \,(k_{n}-i k_{y})\, (q_{m}+i k_{y}) \sqrt{\chi_{m}^3 (k_{z}+\chi_{m})}}\,,\nn
 {u}_{34}^{nm} & = \frac{ (k_{n}+q_{m})\, (q_{m}-i k_{y})\, e^{-\frac{i q_{m} L}{2}}
  \sqrt{\chi_{n}^3 (k_{z}+\chi_{n})} \left [
  k_{n} (k_{z}-\chi_{m})+q_{m} (\chi_{n}-k_{z})+i k_{y} (\chi_{n}-\chi_{m})
  \right ] J_{n-m}\left( \frac{V_{1}}{\hbar \omega}\right)}
  {4\, k_{n}\, \chi_{n} \,(k_{n}-i k_{y})\, (q_{m}+i k_{y}) \sqrt{\chi_{m}^3 (\chi_{m}-k_{z})}}\,,
 \end{align}
 \begin{align}
 {u}_{41}^{nm} & = \frac{ (q_{m}-k_{n})\, (q_{m}+i k_{y})
 \, e^{\frac{i q_{m} L}{2}} \sqrt{\chi_{n}^3 (\chi_{n}-k_{z})} 
 \left[ k_{n} (k_{z}+\chi_{m})+q_{m} (k_{z}+\chi_{n})-i k_{y} (\chi_{n}-\chi_{m})
 \right ] J_{n-m}\left( \frac{V_{1}}{\hbar \omega}\right)}
 {4\, k_{n}\, \chi_{n} \,(k_{n}-i k_{y})\, (q_{m}-i k_{y}) \sqrt{\chi_{m}^3 (k_{z}+\chi_{m})}}\,,\nn
 {u}_{42}^{nm}& = \frac{ (q_{m}-k_{n})\, (q_{m}+i k_{y}) \,e^{\frac{i q_{m} L}{2}} \sqrt{\chi_{n}^3 (\chi_{n}-k_{z})} \left[ k_{n} (k_{z}-\chi_{m})+q_{m} (k_{z}+\chi_{n})-i k_{y} (\chi_{n}+\chi_{m})
 \right ] J_{n-m}\left( \frac{V_{1}}{\hbar \omega}\right)}
 {4\, k_{n}\, \chi_{n} \,(k_{n}-i k_{y})\, (q_{m}-i k_{y}) \sqrt{\chi_{m}^3 (\chi_{m}-k_{z})}}\,,\nn
 {u}_{43}^{nm}& = \frac{ (k_{n}+q_{m})\, (q_{m}-i k_{y}) \,
 e^{-\frac{i q_{m} L}{2}} \sqrt{\chi_{n}^3 (\chi_{n}-k_{z})} 
 \left[-k_{n} (k_{z}+\chi_{m})+q_{m} (k_{z}+\chi_{n})+i k_{y} (\chi_{n}-\chi_{m}) \right ]
 J_{n-m}\left( \frac{V_{1}}{\hbar \omega}\right) }
 {4\, k_{n}\, \chi_{n} \,(k_{n}-i k_{y})\, (q_{m}+i k_{y}) \sqrt{\chi_{m}^3 (k_{z}+\chi_{m})}}
\,,\nn 
 {u}_{44}^{nm}& = \frac{ (k_{n}+q_{m})\, (q_{m}-i k_{y})\,
  e^{-\frac{i q_{m} L}{2}} \sqrt{\chi_{n}^3 (\chi_{n}-k_{z})} 
 \left[  k_{n} (\chi_{m}-k_{z})+q_{m} (k_{z}+\chi_{n})+i k_{y} (\chi_{n}+\chi_{m}) \right [
J_{n-m}\left( \frac{V_{1}}{\hbar \omega}\right)  }
  {4\, k_{n}\, \chi_{n} \,(k_{n}-i k_{y})\, (q_{m}+i k_{y}) \sqrt{\chi_{m}^3 (\chi_{m}-k_{z})}} \,,
  \\
  {M}_{r} ^{nm}& =  e^{-\frac{ i k_{n} L}{2}} \delta_{n,m}\,.
 \end{align}
 
In terms of the above matrices, we can write the wave-function matching relations as:
\begin{align}
A_{n,1}^{i} \,e^{-\frac{ i k_{n}L} {2}} &=  \sum_{m}
 \left (\alpha_{m,1} \,u_{11}^{nm}+\alpha_{m,2} \,u_{12}^{nm}
 +\beta_{m,1} \,u_{13}^{nm}+\beta_{m,2} \,u_{14}^{nm} \right),\nn
 A_{n,2}^{i}\,e^{-\frac{ i k_{n}L} {2}} & =  \sum_{m}
 \left( \alpha_{m,1} \,u_{21}^{nm}+\alpha_{m,2} \,u_{22}^{nm}
 +\beta_{m,1}\, u_{23}^{nm}+\beta_{m,2} \,u_{24}^{nm} \right),\nn
A_{n,1}^{o} & =  \sum_{m}\left ( \alpha_{m,1} v_{11}^{nm}+\alpha_{m,2} v_{12}^{nm}+\beta_{m,1} \,v_{13}^{nm}+\beta_{m,2}\, v_{14}^{nm} \right)
,\nn
A_{n,2}^{o} & =  \sum_{m} \left (
\alpha_{m,1} \,v_{21}^{nm}+\alpha_{m,2}\,
 v_{22}^{nm}+\beta_{m,1} \,v_{23}^{nm}+\beta_{m,2}\, v_{24}^{nm}
\right),
\end{align}
\begin{align}
B_{n,1}^{i}\,e^{-\frac{ i k_{n}L} {2}} & =  \sum_{m}
 \left( \alpha_{m,1} u_{31}^{nm}+\alpha_{m,2} 
 \,u_{32}^{nm}+\beta_{m,1} u_{33}^{nm}+\beta_{m,2} u_{34}^{nm} \right),\nn
 B_{n,2}^{i}\,e^{-\frac{ i k_{n}L} {2}} & =  \sum_{m}
\left( \alpha_{m,1}\, u_{41}^{nm}+\alpha_{m,2} u_{42}^{nm}+\beta_{m,1} 
\,u_{43}^{nm}+\beta_{m,2} \,u_{44}^{nm} \right),\nn
B_{n,1}^{o} & =  \sum_{m}\left(
\alpha_{m,1} \,v_{31}^{nm}
+\alpha_{m,2}\, v_{32}^{nm}+\beta_{m,1}\, v_{33}^{nm}+\beta_{m,2} \,v_{34}^{nm}
\right),\nn
B_{n,2}^{o} & =  \sum_{m}
\left(\alpha_{m,1} \,v_{41}^{nm}+\alpha_{m,2} \,v_{42}^{nm}+\beta_{m,1} v_{43}^{nm}
+\beta_{m,2} \,v_{44}^{nm} \right).
 \end{align}  
 In the matrix form, we can rewrite the above as:

\begin{align}
 {A}_{1}^{o} &=   {v}_{11} \cdot{\alpha}_{1}
 +  {v}_{12} \cdot {\alpha}_{2}+  {v}_{13} \cdot {\beta}_{1}+  {v}_{14} \cdot {\beta}_{2} \,,\quad
 {A}_{2}^{o}  =   {v}_{21} \cdot {\alpha}_{1}+  {v}_{22} \cdot {\alpha}_{2}+ {v}_{23}
  \cdot {\beta}_{1}+  {v}_{24} \cdot {\beta}_{2} \,,\nn
 {B}_{1}^{o} &=    {v}_{31} \cdot {\alpha}_{1}+ {v}_{32} \cdot{\alpha}_{2}+
  {v}_{33} \cdot {\beta}_{1}+ {v}_{34} \cdot {\beta}_{2}\,,\quad
{B}_{2}^{o} =     {v}_{41} \cdot {\alpha}_{1}+
 {v}_{42} \cdot {\alpha}_{2}+  {v}_{43} \cdot 
 {\beta}_{1}+ {v}_{44} \cdot {\beta}_{2}\,,
\end{align}
and
 \begin{align}
 \label{106}
& {A}_{1}^{i}\cdot {M}_{r}= {u}_{11} \cdot {\alpha_{1}}+{u}_{12} \cdot {\alpha_{2}}+{u}_{13} \cdot {\beta_{1}}+{u}_{14} \cdot {\beta_{2}}\,,\quad 
 {A}_{2}^{i}\cdot {M}_{r}= {u}_{21} \cdot {\alpha_{1}}+{u}_{22} \cdot {\alpha_{2}}+{u}_{23} \cdot {\beta_{1}}+{u}_{24} \cdot {\beta_{2}}\,,\nn
& {B}_{1}^{i}\cdot {M}_{r}= {u}_{31} \cdot {\alpha_{1}}+{u}_{32} \cdot {\alpha_{2}}+{u}_{33} \cdot {\beta_{1}}+{u}_{34} \cdot {\beta_{2}}\,,\quad
{B}_{2}^{i}\cdot {M}_{r}= {u}_{41} \cdot {\alpha_{1}}+{u}_{42} \cdot {\alpha_{2}}+{u}_{43} \cdot {\beta_{1}}+{u}_{44} \cdot {\beta_{2}}\,.
 \end{align}
 
 Now, we define a bigger matrix $mat$ as follows:
 \begin{align}
{mat}=
 \begin{pmatrix}
 {u}_{11} & {u}_{12}&{u}_{13}&{u}_{14}  \\
 {u}_{21} & {u}_{22}&{u}_{23}&{u}_{24}   \\
 {u}_{31} & {u}_{32}&{u}_{33}&{u}_{34}  \\
 {u}_{41} & {u}_{42}&{u}_{43}&{u}_{44}   \\
 \end{pmatrix}\,,
  \end{align}  
and denote its inverse as:
    \begin{align}
 {mat}^{-1}=
 \begin{pmatrix}
 {\gamma}_{11} & {\gamma}_{12}&{\gamma}_{13}&{\gamma}_{14}  \\
 {\gamma}_{21} & {\gamma}_{22}&{\gamma}_{23}&{\gamma}_{24}   \\
 {\gamma}_{31} & {\gamma}_{32}&{\gamma}_{33}&{\gamma}_{34}  \\
 {\gamma}_{41} & {\gamma}_{42}&{\gamma}_{43}&{\gamma}_{44}   \\
 \end{pmatrix}\,.
 \end{align}
 Then, we can finally express the amplitudes as:
    \begin{align}
\begin{pmatrix}
{A}_{1}^{o}  \\
{A}_{2}^{o}   \\
{B}_{1}^{o} \\
{B}_{2}^{o}   \\
\end{pmatrix}=
 \begin{pmatrix}
 {M}_{11} & {M}_{12}&{M}_{13}&{M}_{14}  \\
 {M}_{21} & {M}_{22}&{M}_{23}&{M}_{24}   \\
 {M}_{31} & {M}_{32}&{M}_{33}&{M}_{34}  \\
 {M}_{41} & {M}_{42}&{M}_{43}&{M}_{44}   \\
 \end{pmatrix}\begin{pmatrix}
 {A}_{1}^{i}  \\
 {A}_{2}^{i}   \\
 {B}_{1}^{i} \\
 {B}_{2}^{i}   \\
 \end{pmatrix},
 \end{align} 
  where 
\begin{align}
  &{M}_{ij}= \sum_{l=1}^{4}({v}_{il} \cdot {\gamma}_{lj})\cdot {M}_{r}\,.
  \end{align} 
For the $n^{\text{th}}$ Floquet band, the above matrix equation reduces to: 
\begin{align}
\begin{pmatrix}
{A}_{1n}^{o}  \\
{A}_{2n}^{o}   \\
{B}_{1n}^{o} \\
{B}_{2n}^{o}   \\
\end{pmatrix}=
\sum_{m=-\infty}^\infty\mathcal{S}_{nm}\begin{pmatrix}
{A}_{1m}^{i}  \\
{A}_{2m}^{i}   \\
{B}_{1m}^{i} \\
{B}_{2m}^{i}   \\
\end{pmatrix}\,,
\end{align}  
where we can determine $\mathcal{S}_{nm}$ from ${M}_{ij}$.

\end{document}